\documentclass{article}
\usepackage[utf8]{inputenc}
\usepackage{subcaption} 
\usepackage{booktabs}
\usepackage[english]{babel}
\usepackage{multicol}
\usepackage{nicematrix}
\usepackage{indentfirst}
\usepackage{multirow}
\usepackage{amsmath, soul}
\usepackage{caption}
\usepackage[breaklinks=true]{hyperref}
\usepackage{xurl}
\usepackage[section]{placeins}
\usepackage{bm,graphicx}
\usepackage{float}
\usepackage{tikz}
\usepackage{amsmath}
\usepackage{tikz}
\usetikzlibrary{arrows.meta}
\usetikzlibrary{tikzmark}
\usepackage{enumitem,caption}

\usepackage[letterpaper,margin=1in]{geometry}
\usepackage[backend=biber, style=numeric, sorting=none, giveninits=true, maxnames=19, minnames=1, maxbibnames=19, minbibnames=1]{biblatex}
\newlength{\symbolwidth}
\usepackage[
singlelinecheck=false 
]{caption}
\usepackage{color,soul}

\begin{document}

\title{Predictability and Statistical Memory \\ in Classical Sonatas and Quartets}
\author{Linus Chen-Plotkin,$^{1,\,2}$ Suman S. Kulkarni,$^{2,\,3}$ and Dani S. Bassett$^{2,\,3,\,4,\,5,\,6,\,7}$}
\date{%
    $^1$\textit{Germantown Friends School, Philadelphia, PA 19144, USA}\\%
    $^2$\textit{Department of Physics \& Astronomy, College of Arts \& Sciences,
University of Pennsylvania, Philadelphia, PA 19104, USA}\\%
    $^3$\textit{Department of Bioengineering, School of Engineering \& Applied Science,
University of Pennsylvania, Philadelphia, PA 19104, USA}\\%
    $^4$\textit{Department of Electrical \& Systems Engineering, School of Engineering \& Applied Science, University of Pennsylvania, Philadelphia, PA 19104, USA}\\%
    $^5$\textit{Department of Neurology, Perelman School of Medicine,
University of Pennsylvania, Philadelphia, PA 19104, USA}\\%
    $^6$\textit{Department of Psychiatry, Perelman School of Medicine,
University of Pennsylvania, Philadelphia, PA 19104, USA}\\%
    $^7$\textit{Santa Fe Institute, Santa Fe, NM 87501, USA}\\[2ex]%
    (Dated: September 28, 2025)
}

\maketitle
\begin{abstract}     
\noindent Statistical models and information theory have provided a useful set of tools for studying music from a quantitative perspective. These approaches have been employed to generate compositions, analyze structural patterns, and model cognitive processes that underlie musical perception. A common framework used in such studies is a Markov chain model, which models the probability of a musical event---such as a note, chord, or rhythm---based on a sequence of preceding events. While many studies focus on first-order models, relatively few have used more complex models to systematically compare across composers and compositional forms. In this study, we examine statistical dependencies in classical sonatas and quartets using higher-order Markov chains fit to sequences of top notes. Our data set of 605 MIDI files comprises piano sonatas and string quartets by Wolfgang Amadeus Mozart (1756–1791), Joseph Haydn (1732–1809), Ludwig van Beethoven (1770–1827), and Franz Schubert (1797–1828), from which we analyze sequences of top notes. We probe statistical dependencies using three distinct methods: Markov chain fits, time-delayed mutual information, and mixture transition distribution analysis. We find that, in general, the statistical dependencies in Mozart's music notably differ from that of the other three composers. Markov chain models of higher order provide significantly better fits than low-order models for Beethoven, Haydn, and Schubert, but not for Mozart. At the same time, we observe nuances across compositional forms and composers: for example, in the string quartets, certain metrics yield comparable results for Mozart and Beethoven. Broadly, our study extends the analysis of statistical dependencies in music, and it highlights systematic distinctions in the predictability of sonatas and quartets from different classical composers. These findings motivate future work comparing across composers for other musical forms, or in other eras, cultures, or musical traditions. 
\end{abstract}
     
\section{Introduction}

Since the publication of Claude E. Shannon's ``A Mathematical Theory of Communication" in 1948 \cite{shannon1948mathematical}, a variety of concepts and techniques from information theory have been used to understand the structure of music. In 1957, music theorist Leonard B. Meyer argued that entropy may be considered as a rough metric of value and meaning in music \cite{meyer1957meaning}. Since then, more nuanced theories of expectation, perception, and enjoyment have been put forth \cite{thompson2023psychological,huron2008sweet,stupacher2022sweet,cheung2019uncertainty,koelsch2010}. A large body of work has since used information theory and probabilistic models to examine the structure of musical compositions and music perception \cite{pearce20012,pearce2003,pearce2004}. Prior work has modeled melodic and harmonic structure using probabilistic methods \cite{allan2005harmonising,roig2014knowledge,arthur2017taking}. In music cognition, variations of Markov chains are used to model listener perception as a dynamic process by updating transition weights with each new note \cite{ cheung2019uncertainty, temperley2008probabilistic,abdallah2009information}. Markov models have also been used in combination with generative algorithms in order to identify the genre or composer of a score, or to generate novel scores with desired characteristics \cite{kaliakatsos2011weighted,van2010music}. Approaches from network theory have also provided new ways to study musical form, harmony, and complexity \cite{ferretti2017modeling, buongiorno2020topology,mrad2024network}. In combination with information theory, these approaches have been used to model how humans perceive the information content of musical compositions \cite{kulkarni2024information,lynn2020human}. Additionally, recent developments have emphasized the role of higher-order statistical structure in music, for example through topology-based analyses of Bach’s works \cite{mrad2025higher} or multivariate information measures such as O-information \cite{O-information}.

Although the statistical structure of music has been studied extensively, fewer studies compare composers within a shared form while jointly assessing model order and multiple dependency metrics. Hidden Markov Models have been employed to attribute melodies to composers \cite{pollastri2001classification}, but these models do not reveal which musical features differentiate one composer from another, since the underlying states are hidden. Low-order Markov chains have also been used to classify melodies by Mozart, Beethoven, and Haydn \cite{kaliakatsos2011weighted,liu2002modeling}, though these studies emphasized classification performance rather than systematic comparisons of the underlying dependency structures. Given that these models are successful at distinguishing between compositional forms and composers, it seems reasonable that there may be quantifiable aspects of musical style that vary between them. Here we address this gap by comparing statistical dependencies across composers and forms by measuring them in three distinct ways: time-delayed mutual information, first- vs. second-order Markov models, and mixture transition distributions. We focus on musical predictability, often linked to musical appreciation \cite{pearce2012auditory}, and quantify predictability using statistical dependencies---the timescale on which musical events influence future ones in a composition. This motivates the main question we focus on in this work: How do statistical dependencies differ across classical composers, and what informational properties of the music account for those differences?

In this study, we measure statistical dependencies in musical data using three distinct approaches---Markov chains of first and second order, time-delayed mutual information, and mixture transition distribution modeling (MTD)\cite{berchtold2002mixture}---to identify specific differences between the styles of Mozart, Beethoven, Haydn, and Schubert. Each of these methods evaluates statistical dependencies, but quantify different aspects: mutual information quantifies the strength of dependencies at different lags; Markov-chains can be used to determine which order $n$-gram best describes the data; and MTD models approximate higher-order Markov processes by assuming that the influence of multiple past notes is additive, with weights that indicate the relative contribution of each lag. We consider piano sonatas and string quartets due to their length, established compositional norms, and formal relations \cite{rosen1980sonata, bigo2017sketching}, and we focus our analysis on the melody line. In these melodic lines, we then investigate how \textit{recent} the melodic information is that predicts a composer's next note choice is, and hence the extent of statistical dependencies. We use statistical tests (particularly the Bayesian information criterion \cite{schwarz1978estimating}) to compare the performances of lower-order models to higher-order models fit to melodies of different composers. By implementing this approach, we determine how predictable a given composer's music is, and the extent of statistical dependencies present in the pieces. 

We collate 316 piano sonatas and 289 string quartets. For each piece, we extract the melody line by selecting the top-notes with the Python library \texttt{Music21}. We apply multiple metrics designed to assess statistical dependencies and long-range correlations in musical sequences, though each emphasizes slightly different aspects of structure. Across these metrics, we observe differences that depend both on the compositional form (sonatas versus quartets) and on the individual composer. We observe that higher-order models---such as second-order Markov chains and mixture transition distribution (MTD) models of order two or three---in general provide better fits for Beethoven, Haydn, and Schubert, whereas Mozart's compositions tend to rely more on immediate note-to-note transitions. Yet, we also observe nuances across compositional forms and composers. For example, for string quartets, we observe that some of Beethoven's compositions are also better explained by models that rely on the immediate preceding note, while for sonatas, the MTD models indicate comparable short-range structure for Haydn and Mozart. Our findings offer a formal account of differences in statistical dependencies across classical composers, and identify informational factors that explain those differences. More broadly, they motivate future work to determine the relation between note-to-note predictability and its consequences for how humans perceive musical aesthetics and complexity.
 
\section{Methods}
\subsection{Data preparation}
\label{prep}

We compiled our dataset from MIDI files available at \url{https://www.kunstderfuge.com}. Each MIDI file tabulates the time, duration, pitch, and volume of each note in a piece. We extracted all piano sonatas of Beethoven, Haydn, Mozart, and Schubert and all string quartets of Beethoven, Haydn, and Mozart. We removed duplicates and omitted files with fewer than a $100$ notes from our dataset. In cases where individual MIDI files were not available for each movement, we manually cut full sonatas into their movements using MuseScore. There are 316 files in the sonata group, each consisting of one movement---63 files for Beethoven, 161 files for Haydn, 66 files for Schubert, and 45 files for Mozart---and 289 files in the string quartet group -- 100 files for Beethoven, 147 files for Haydn, and 35 files for Mozart.

\begin{figure}[htbp]
\begin{center}
    \begin{subfigure}{\linewidth}
    \caption{Reading the MIDI file}
    \centering
        \includegraphics[width=0.5\linewidth]{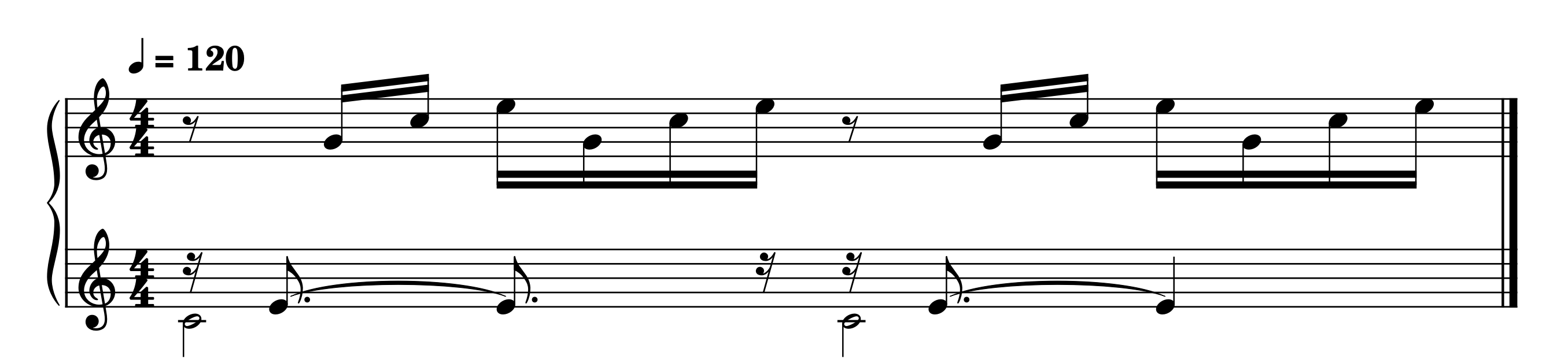}
    \end{subfigure}
    \vspace{0.5em}
    \begin{subfigure}{\linewidth}
        \caption{Chordifying the score}
        \centering
        \includegraphics[width=0.5\linewidth]{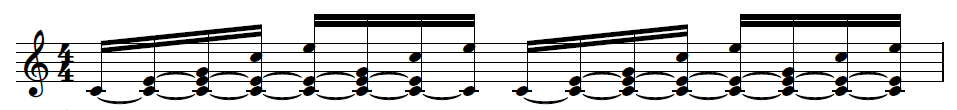}
    \end{subfigure}
    \vspace{0.5em}
    \begin{subfigure}{\linewidth}
        \caption{Selecting the top note and mapping to one octave}
        \centering
        \includegraphics[width=0.5\linewidth]{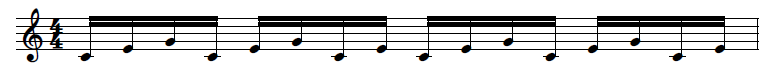}

    \end{subfigure}
    \caption{\textbf{Data extraction and preparation.} (A) Using the python library Music21, we read each MIDI file as a Music21 score object that includes the onset times, durations, and pitches of each note in the song. Pitches are numbered 0-127. By convention, pitch \#60 is middle C, and a MIDI pitch increment of one is equal to a single half-step on an equally tempered scale. (B) We apply Music21's ``chordify" function, which condenses the MIDI streams into a series of chords which contains the pitches of each note that is played at each point in time. (C) From the chordified stream, we extract a monophonic melody by selecting the highest pitch active at each chord onset. We use Music21's \texttt{score.analyze(`key')} function to determine the overall key of the piece and then transpose all pitches in this sequence to a single octave in the key of C. Note that the \textit{pitch class} of a pitch is the set of all pitches that are a whole number of octaves away. For brevity, in the remainder of this manuscript, when we refer to a ``note", we mean the entire pitch class.}
    \label{figmethod}
\end{center}
\end{figure}

Each of these musical compositions are temporal sequences of categorical data. Here we consider monophonic sequences of notes, and the composition's melody specifically, consistent with prior work \cite{pearce2003,pearce2004,pearce20012,eorola2009,Eerola2003}. Our approach, starting with polyphonic files and extracting melody lines differs from earlier studies that used existing datasets of monophonic melodies \cite{pearce2004,hamilton2024billboard}. To extract single-note sequences from each file (see Fig.~\ref{figmethod}), we first ``chordify" the score (Figure \ref{figmethod}(b), then extract a sequence of top notes \cite{kaliakatsos2011weighted} (see Figure \ref{figmethod}(c)) . This method has been used previously to effectively find melody-lines in several empirical datasets \cite{uitdenbogerd1999melodic,kaliakatsos2011weighted}. After extracting the top-note sequence, we standardized pitch representations by transposing each piece into the key of C and collapsing all pitches into a single octave. This step ensures that analyses are based on relative pitch classes rather than the absolute register, so that notes separated by octaves are treated as equivalent. Throughout the manuscript, when we refer to a ``note", we specifically mean a pitch class.

\subsection{Choice of Corpus}

Each of the composers we study possesses unique qualities as well as shared features in the classical style \cite{rosen1971classical}. Joseph Haydn (1732-1809) and Wolfgang Amadeus Mozart (1756-1791) are both classical composers. Ludwig van Beethoven (1770-1827) wrote much of his work during the classical period, but some of his late work may be considered early romantic. Franz Peter Schubert (1797-1828) is considered a composer of the romantic period, and overlaps with the later part of Beethoven's life. Among these composers, we choose to focus on sonatas and string quartets because they are lengthy musical forms (amenable to statistical analysis) that have established (and intertwined) norms \cite{rosen1980sonata,bigo2017sketching}. The sonata form, used both in piano sonatas and in string quartets, dictates that the main musical theme passes through development, exposition, and recapitulation \cite{benward2009music}. Because of the shared structure of the the sonata and quartet forms, any observed differences among composers would reflect differences in implementation rather than formal constraint. 

\subsection{Quantifying Statistical Dependencies}

To quantify statistical dependencies in these pieces, we use three distinct approaches. First, we utilize time-delayed mutual information to quantify the strength of dependencies between notes that are separated by different numbers of intervening notes. Second, we fit Markov chain models of different orders to compare the predictability of sequences of two consecutive notes to sequences of three consecutive notes in a composition. Third, we apply a mixture transition distribution model to extend our results from Markov Chains to higher orders. Together, these approaches allow us to build an understanding of predictability in the piano sonatas and string quartets of Haydn, Mozart, Beethoven, and Schubert.

\subsubsection{Time-delayed Mutual Information}

Mutual information quantifies the statistical dependency between two random variables by measuring how much knowing one reduces uncertainty about the other~\cite{cover1999elements,steuer2002mutual}. The approach has been used to study the coding of music in the brain \cite{kim2021dissociation}, as well as more generally in the study of human physiology \cite{albers2012using}, molecular biology \cite{yang2016inference} and cybersecurity ~\cite{ma2025cyber}. To begin, let $X$ and $Y$ be two jointly discrete random variables with state spaces $\mathcal{X}$ and $\mathcal{Y}$, respectively. The mutual information between the two quantities is defined as:
\[I(X;Y) = \sum_{y\in\mathcal{Y}} \sum_{x\in\mathcal{X}} P_{(X,Y)}(x,y)\log\left(\frac{P_{(X,Y)}(x,y)}{P_X(x)P_Y(y)}\right) ,\] 
where $P_{(X,Y)}$ denotes the joint probability mass function of $X$ and $Y$, and $P_X$ and $P_Y$ are the marginal probability mass functions of $X$ and $Y$, respectively. Mutual information between two variables $X$ and $Y$ can also be expressed in terms of the marginal entropies $H(X)$ and $H(Y)$ and the conditional entropies $H(X|Y)$ and $H(Y|X)$:
\begin{align*}
    I(X;Y) &= H(X) - H(X|Y)\\
    &= H(Y) - H(Y|X) .
\end{align*}
Throughout, we compute mutual information using logarithm base $2$, and hence results are reported in bits such that $I(X,Y)$ reflects how many bits of information we learn about $Y$ through knowing $X$, or vice versa. A value of $0$ for the mutual information indicates that $X$ and $Y$ are independent, and a mutual information above $0$ indicates some degree of dependency between $X$ and $Y$.

To study sequential dependencies in musical structure, we compute the time-delayed mutual information between notes separated by a lag $\tau$. Conceptually, the mutual information between notes separated by a given lag captures how much knowing the current note reduces uncertainty about the note $\tau$ steps later. Given a time series of length $N$ with note values ranging from 1 to 12 extracted from a MIDI file as described in Section \ref{prep}) and a lag $\tau$, we use mutual information to quantify dependencies between notes that are $\tau$ time steps apart. Let $X$ be an arbitrary note in the series that falls before time $N-\tau$, and let $Y$ be the note that occurs $\tau$ time-steps later. We compute the joint probability mass function $P(X,Y)$ by counting the number of times that $x_i$ occurs in the time series and $y_j$ occurs $\tau$ steps later, for each $x_i$ and $y_j$ in the state space. The marginal probability functions $P(X)$ and $P(Y)$ are found by counting as well. From here, we then quantify the mutual information at lag $\tau$ as:

\begin{equation}\label{eq:mutual_info}
    I(\tau) = I(X(t), X(t + \tau)) = \sum_{x_t} \sum_{x_{t + \tau}} P(x_t, x_{t + \tau}) \log \left( \frac{P(x_t, x_{t + \tau})}{P(x_t) P(x_{t + \tau})} \right) .
\end{equation}
We use the above quantity to characterize the predictability of musical sequences across temporal scales.

\subsubsection{Markov Chains and Bayesian Information Criterion}

A second approach to quantifying sequential dependencies is to model note sequences as Markov chains and evaluate their fit using likelihood-based criteria such as the Bayesian Information Criterion (BIC) \cite{schwarz1978estimating}. A $k$-th order Markov chain is a random process in which the probability of moving to the next state depends only on the previous $k$ states \cite{lamperti1977stochastic}. Formally, a stochastic process ${X_t}$ satisfies the Markov property of order $k$ if
\begin{equation}\label{eq:markov_k}
    \Pr (X_t = x | X_{t-1}, X_{t-2}, ..., X_0) = \Pr(X_t = x| X_{t-1}, ... , X_{t-k}).
\end{equation}

In our case, states correspond to melodic notes, and the order of the chain determines how many preceding notes are taken into account when predicting the next one. For example, in a first-order Markov chain, the next note depends only on the immediately preceding note, while in a second-order chain it depends on the two preceding notes. 

This approach provides a complementary perspective to the mutual information approach. In the mutual information approach, we note that lagged dependencies are assessed independent of each other. For instance, when computing the mutual information at lag two, we ignore the middle note in a sequence of three notes. By contrast, a Markov-2 model explicitly parameterizes the dependencies across all three notes. As a result, we expect a Markov-2 model to be sensitive to information carried by musical devices---such as a recurrent three note motif, for instance---that may not be captured by mutual information alone. Higher-order Markov chains hence provide information about predictability in a distinct way.

In practice, we restrict our attention to first- and second-order models. This is due to the high number of free parameters in Markov-3 model (and above), relative to the limited amount of data in each musical piece. A Markov-3 model would have 19008 free parameters, which is more than the number of notes in most of the pieces in our datasets.

\subsubsection{Fitting first and second order Markov chains}

We now describe how we fit discrete-time, discrete-state Markov chains to sequences of notes and the methods we use to assess goodness of fit. A first-order Markov chain is a stochastic process characterized by the property that the conditional probability of a given state $x$ occurring at time $t$ in the process depends only upon the previous state $X_{t-1}$. That is:
\begin{equation*}
    \Pr(X_t = x \mid X_{t-1} = x_{t-1}, X_{t-2} = x_{t-2},...,X_0 = x_0) = Pr(X_t=x \mid X_{t-1} =x_{t-1}).
\end{equation*}
A Markov-1 chain with a state space of $m$ unique states can be parameterized by an $m\times m$ transition matrix $P$, where each entry $p_{ij}$ gives the probability of moving from state $x_i$ to state $x_j$.
\begin{equation}
P = 
\begin{NiceMatrix}[baseline=4]
X_{t-1}& & X_t & \\
& 1 & \dots & m\\
1 & p_{11} & \dots & p_{1m}\\
\vdots & \vdots & \ddots\\
m & p_{m1} & \dots  & p_{mm}  
\CodeAfter \SubMatrix[{3-2}{5-4}]
\end{NiceMatrix},
\end{equation}
where $p_{ij} = \Pr(X_{t} = x_j \mid X_{t-1} = x_i)$. Also note that by probability conservation, we have $\sum\limits_{j=1}^m p_{ij} = 1$ for all $i$. The maximum-likelihood estimate of $p_{ij}$ for a sequence $S$ is obtained by counting the transitions:
\begin{equation}
    p_{ij} = \frac{n_{ij}}{\sum_{k \in S} n_{ik}},
\end{equation}
where $n_{ij}$ is the number of times that state $j$ immediately follows state $i$ in $S$.

More generally, as described in Eq.~\ref{eq:markov_k}, a Markov chain of arbitrary order $k$ is characterized by the property that the conditional probability of a given state occurring at time $t$ depends only upon the past $k$ states. A second-order Markov chain therefore captures the joint influence of two preceding states. The transition probabilities can be written as
\begin{equation}
    p_{ijk} = Pr(X_t = k \mid X_{t-1} = j, X_{t-2} = i)
\end{equation}
and collected into a transition probability matrix of dimension $m^k \times m$, where each row corresponds to a two-note tuple $(i,j)$ and each entry specifies the probability of transitioning into a third note $k$:
\[P_2 = 
\begin{NiceMatrix}[baseline=4]
\Block{1-1}{X_{t-2}} & X_{t-1} &  & X_t &\\
& & 1 & \dots & m\\
1 & 1 & _1p_{11}  & \dots & _1p_{1m}\\
1 & \vdots & \vdots & \ddots & \vdots\\
1 & m & _1p_{m1} & \dots  & _1p_{mm}\\
2 & 1 &  _2p_{11} & \dots & _2p_{1m}\\
2 & \vdots & \vdots & \ddots & \vdots\\
2 & m & _2p_{m1} & \dots  & _2p_{mm}\\
\vdots & \vdots & \vdots & \ddots & \vdots\\
m & m & _mp_{m1} & \dots  & _mp_{mm}\\
\CodeAfter \SubMatrix[{3-3}{10-5}]
\end{NiceMatrix}\]
If we are given $m-1$ of the entries in a row, we can determine the value of the other entry because we know the row must sum to one. Therefore the maximal number of free parameters in a Markov-$k$ chain on a space of $m$ states is $m^k(m-1)$.

In our work, we fit first- and second-order Markov chains to sequences of notes extracted from MIDI files as described in Section \ref{prep}. The state space of these sequences are the pitch classes of each note in the chromatic scale, which can be represented by $\{C,C\#,D,D\#,E,F,F\#,G,G\#,A,A\#,B\}$ or any enharmonically equivalent set of notes.

\subsubsection{Likelihood ratio and BIC tests for Markov chains}\label{LLR_BIC_methods}

To quantify how well a probabilistic Markov model fits the data, we compute the likelihood of a sequence occurring under the model. For a first-order chain with transition probability matrix $P$, the likelihood of a sequence $S = \{S_1, S_2, ..., S_n \}$ is
\begin{equation}
    L = \prod_{i=1}^{n-1} P(S_i, S_{i+1}).
\end{equation}
In practice, we work with the log-likelihood to avoid numerical inaccuracies when dealing with extremely small floats:
\begin{equation}
    \log(L) = \log\left( \prod_{i=1}^{n-1} P(S_i, S_{i+1})\right) = \sum_{i=1}^{n-1} \log \left( P(S_i, S_{i+1}) \right).
\end{equation}
For a Markov model of order $k$, the log-likelihood generalizes to
\begin{equation}
    \log(L) = \sum_{i=1}^{n-l} \log \left(P(S_i,S_{i+1},...,S_{i+l}) \right).
\end{equation}

The likelihood (or log-likelihood) of a piece will naturally increase as the number of parameters used in the Markov model. And so likelihood alone cannot be used to compare models of different order. To test whether the difference in log-likelihood between a first-order model and a second-order model is large enough to reject the Markov-1 process as an effective explanation, we employ a likelihood ratio test \cite{anderson1957statistical}. Assuming the null hypothesis that the data follow a first-order chain, the likelihood ratio statistic is 
\begin{equation}
   LR = -2 \log \left( \frac{L(m_1)}{L(m_2)} \right), 
\end{equation}
where $L(m_1)$ and $L(m_1)$ are the likelihoods under the first- and second-order models, respectively. Comparing our ratio to the chi-squared distribution with degrees of freedom equal to the difference in the maximum number of free parameters between the two models, we obtain a $p$-value. For $m$ states, a Markov-$k$ model has $m^k (m-1)$ free parameters, so with $m=12$ pitch classes, the comparison between orders one and two gives $(12-1) \times 12^2 - (12-1) \times 12 = 1452$ degrees of freedom.

We use this test to assess whether a second-order model provides a significantly better fit than a first-order model, where we apply it to our dataset of sonatas and quartets. However, the likelihood ratio test is known to be biased by the length of a piece: it is more likely to tell us a long piece is generated by Markov-2 than a short piece \cite{bartlett1937properties} (see top row of Supplementary Figure \ref{LL_len}). As a result, the likelihood-ratio test can be skewed according to the average piece length by each composer.

To address this limitation, we also consider the Bayesian Information Criteria (BIC), which adjusts the log-likelihood by introducing a penalty for the number of free parameters and the sequence length:
\begin{equation}
    \mathrm{BIC} = -2 \log(L) + k\log(n),
\end{equation}
where $k$ is the number of free parameters in the model and $n$ is the length of the sequence. For transition matrices, $k$ is taken to be the number of free (non-redundant) entries, following standard convention \cite{raftery1985model}. Models with lower BIC are preferred \cite{katz1981some}.

We compare first- and second-order models by computing
\begin{equation}\label{eq:delta_BIC}
        \Delta \mathrm{BIC} = \mathrm{BIC}(m_1) - \mathrm{BIC}(m_2)  
\end{equation}
A positive $\Delta \mathrm{BIC}$ value indicates that a second-order model provides a better fit than a first-order model after accounting for model complexity and sequence length. Unlike log-likelihood tests, the $\Delta$ BIC values are not correlated with piece length (see bottom row of Supplementary Figure \ref{LL_len}).

\subsubsection{Models and inference from Mixture Transition Distributions}

While likelihood-based tests provide a method to evaluate low-order Markov models, two limitations remain. First, the likelihood ratio test statistic is only asymptotically a chi-squared distribution \cite{wilks1938large}. The test may be less conservative when the number of samples is small compared to the number of parameters \cite{bartlett1937properties}. In our case even a second-order Markov chain involves a considerable number of parameters (up to $1,548$ in our dataset), whereas the musical sequences are often fewer than $2,000$ notes. This makes the likelihood-ratio test less reliable and limits our ability to examine yet higher order models. Second, low-order Markov models provide a limited description of musical structure. 

As a step toward overcoming these limitations, we employ a family of higher-order Markov models called mixture transition distribution (MTD) models \cite{raftery1985model,berchtold2002mixture}. MTDs capture some aspects of higher-order structure while requiring fewer parameters than arbitrary Markov models. Instead of assuming a full transition probability conditioned on the past $k$ states, an MTD model represents the conditional probability of the next state as an additive mixture of transition probabilities, summed over weighted contributions from the past $k$ states. Formally, a $k$-th order MTD assumes that the probability of observing a state $x_0$ at time $t$, given the past $k$ states $\{ X_{t-1}, X_{t-2},...,X_{t-k} \}$ is given by the mixture distribution
\begin{equation}
    P(X_t = x_0 \mid X_{t-1} = x_1, \dots, X_{t-l} = x_k) = \sum_{n=1}^{k} \beta_n q(x_n,x_0),
\end{equation}
where $q$ is a single $m \times m$ transition matrix over the state space, and $\beta_n$ specifies the relative influence of the $n$-th lag on predicting the next note. The weights $\beta_n$ are nonnegative and $\sum_{n=1}^{k} \beta_n = 1$.

For example, in a second-order MTD, the probability of transitioning from $i \rightarrow j \rightarrow k$ is expressed as 
\begin{equation}
    P(X_t = k | X_{t-1} = j, X_{t-2} = i) = \beta_1 q_{jk} + \beta_2 q_{ik}
\end{equation}
Thus, unlike an arbitrary second-order Markov chain with up to $m^2(m-1)$ parameters, the MTD requires only $m(m-1)$ parameters plus $k=2$ mixture weights. For each increase in order $k$, only one additional free parameter is introduced. 

We estimate the parameters of the MTD using an expectation-maximization (EM) algorithm \cite{berchtold2020optimization} that maximizes the likelihood of the observed sequences over both the transition matrix $q$ and the weights $\beta_n$. By examining the fitted weights, we assess the relative importance of each lag. This enables us to obtain insight into the role of deeper memory (higher order lags), while overcoming the limitations of introducing too many additional parameters.

\section{Results}

\subsection{Mutual information} \label{sec:MI_results}
\begin{figure}[hbt!]
    \centering
    \includegraphics[width=0.85\linewidth]{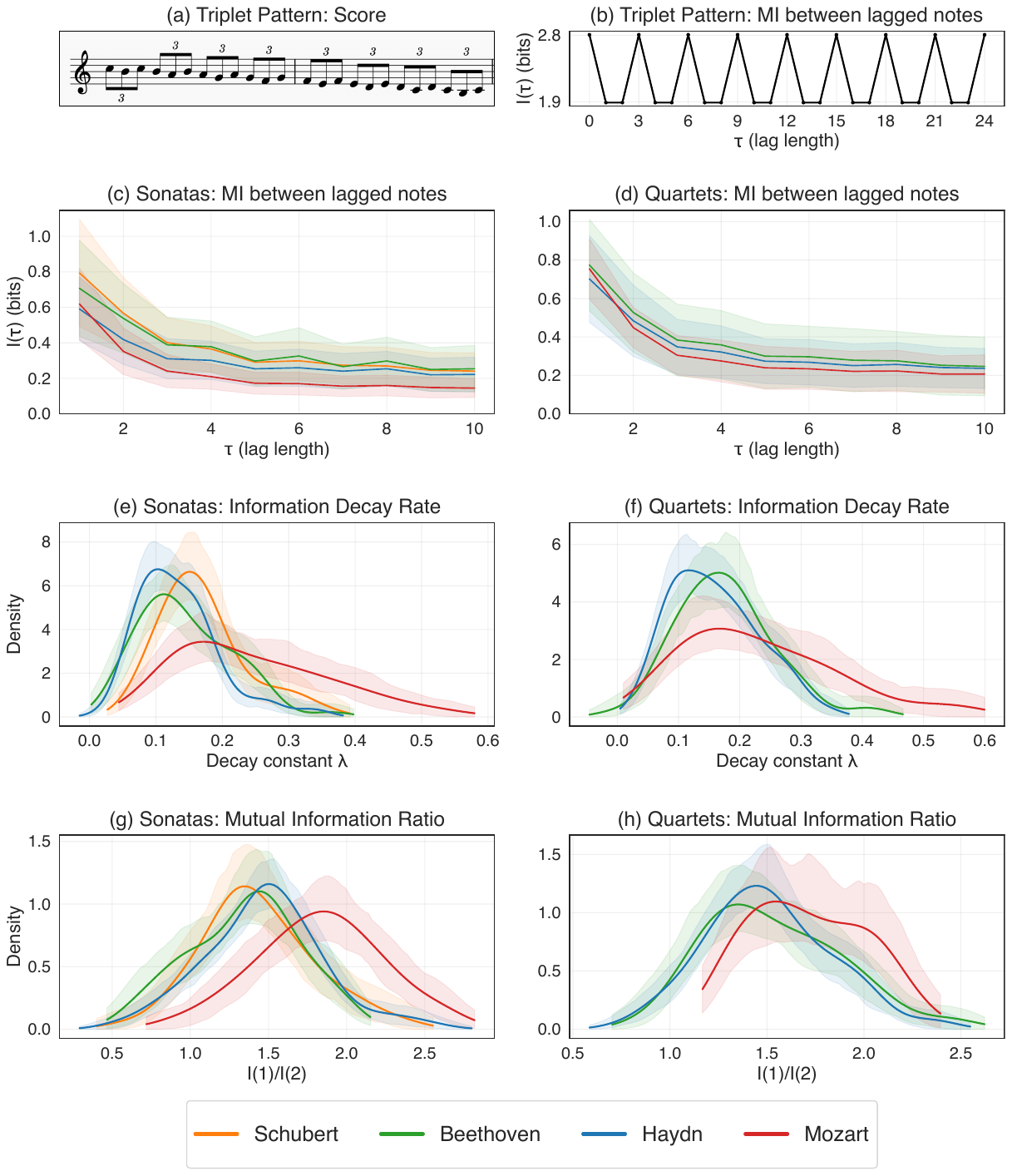}
    \caption{\textbf{Comparative information-theoretic analysis of sonatas and quartets by classical composers.} (a) An example score in which mutual information is maximum between notes at lags that are multiples of three. (b) The mutual information between notes in the example score at lags from 0 to 24. (c) Lagged mutual information \( I(\tau) \) for sonatas across lags \( \tau = 1 \) to \( 10 \), showing how information decays over time for each composer. (d) Same as panel (c), but for string quartets. (e) Kernel density estimates of the decay constant \( \lambda \), measuring the rate of exponential decline in mutual information for sonatas. (f) Decay constants \( \lambda \) for quartets. (g) Distribution of mutual information ratios \( I(1)/I(2) \), representing short-term predictability in sonatas. (h) Mutual information ratios \( I(1)/I(2) \) for quartets.}
    \label{MI_combined_fig}
\end{figure}

To build intuition for how melodic structure can influence mutual information, we first discuss some simple illustrative examples. Consider a line consisting of a continuous ascending scale. Here, every note is determined by any earlier note through a fixed shift. The mutual information between notes at any lag $\tau$ hence reaches its maximum value, and is equal to the entropy of the distribution of notes. A similar argument is true for a trill, in which the melody alternates between two notes, and hence each note can be predicted from any prior note. By contrast, consider a line of descending triplets, as shown in Fig. \ref{MI_combined_fig}(a). Here, predictability (and hence, mutual information) depends on the lag length. Successive notes contain incomplete information about one another because both upwards and downwards movements are possible over a sequence of two consecutive notes, and so the mutual information between adjacent notes is less than maximal. However, notes separated by lags of length three (or six, or nine, etc.) are fully determined by the triplet pattern: sampling every third note yields a descending scale, while sampling every sixth note produces a descending arpeggio. This regularity is reflected in the peaks of mutual information at lags three, six, nine, and so forth, as observed in Fig.~\ref{MI_combined_fig}(b).

In a piece of classical music, we expect the mutual information to vary with lag, usually decreasing with lag since patterns and motifs are temporally localized. To examine this dependency empirically, we compute the mutual information as a function of lag for piano sonatas and string quartets by Mozart, Haydn, Beethoven, and Schubert. We examine lags up to length $10$. As shown in Fig.~\ref{MI_combined_fig}(c) for the sonatas and Fig.~\ref{MI_combined_fig}(d) for the quartets, the mutual information $I(\tau)$ declines as the lag length $\tau$ increases. The rate of this decay reflects how quickly a melody loses predictability: a rapid decay indicates greater unpredictability (or more localized structure) within the piece's melody, whereas a slower decay indicates greater predictability, possibly suggesting recurrent patterns or long unbroken melodic movements.

We quantify this decay for each piece in our dataset by fitting an exponential function to the lagged mutual information
\begin{equation}
    I(\tau) = I_0 \cdot e^{-\lambda \tau},
\end{equation}
where $\lambda$ is the decay constant. This exponential form provides a good description of the decline across lags. For sonatas, the mean $R^2$ values are 0.96 (Mozart), 0.82 (Beethoven), 0.87 (Haydn), and 0.88 (Schubert). For quartets, the mean $R^2$ values are 0.96 (Mozart), 0.95 (Beethoven), and 0.93 (Haydn).

To examine how the decay constants vary across composers, we then plot their distributions in Fig.~\ref{MI_combined_fig}(e) and (f). The summary statistics of these distributions are provided in Supplementary Table.~\ref{sum_decay}. For both sonatas and quartets, we observe that the decay rates ($\lambda$'s) associated with Mozart's compositions tend to be greater than those of other composers. We then formally test whether the average decay constants for Mozart differ from those of the other composers combined. The results of Welch’s $t$-tests are shown in Table~\ref{decay}. In both the sonata dataset ($t$-test, $t$=6.07, $df$ = 48.94, and $p$-value = $1.83\times10^{-7}$) and the quartet dataset ($t$=3.09, $df$ = 34, and $p$-value = $3.78\times10^{-3}$), Mozart’s values are higher than the other composers. These results indicate that the notes in Mozart's sonatas and quartets are not as easily predicted from prior notes, relative to comparable compositions by Beethoven, Schubert, and Haydn. In Supplementary Table ~\ref{pair_lambda}, we also report all pairwise comparisons across composers. 

\begin{table}[h!]
\centering
\caption{\textbf{Group comparisons for the mutual information decay constant.} For each dataset (sonatas and quartets), we report the mean and standard deviation of the fitted decay constants ($\lambda$), along with the number of pieces ($N$) examined for Mozart and for the combined group of other composers. Welch's $t$-tests evaluate whether the average decay constants for Mozart differ significantly from the other composers.}
\label{decay}
\begin{tabular}{llcccc}
\toprule
Dataset & Group & Mean & Std. Dev. & N & $t$-test (Mozart vs Others) \\
\midrule
Sonata & Mozart & 0.2477 & 0.1112 & 45 & \multirow{2}{*}{$t = 6.07,\, p = 1.83\times10^{-7},\, df = 48.94$} \\
       & Others (H+B+S) & 0.1432 & 0.0654 & 278 & \\
\midrule
Quartet & Mozart & 0.2337 & 0.1235 & 34 & \multirow{2}{*}{$t = 3.09,\, p = 0.004,\, df = 37.13$} \\
       & Others (H+B) & 0.1653 & 0.0738 & 194 & \\
\bottomrule
\end{tabular}
\end{table}
We now focus specifically on the shorter time lags $\tau = 1$ and $\tau = 2$. To capture the relative drop in predictability between these two lags, we consider the ratio
\begin{equation}
    \frac{I(1)}{I(2)} ~.
\end{equation} 
This ratio measures how much more information is retained from one step back compared to two steps back. Larger values indicate that predictability falls off more steeply beyond the immediately preceding note.

We measure the ratio of mutual information at lags 1 and 2 for each piece in our dataset. In Fig. \ref{MI_combined_fig}(g)--(h), we show the distribution of this ratio for each composer. The summary statistics of these distributions are provided in Supplementary Table~\ref{sum_ratio}. We observe that Mozart’s pieces are shifted toward higher values relative to the other composers, indicating a steeper decline in predictability between the first and second time lag. We then formally test whether the average ratios for Mozart differ from those of the other composers combined. Table~\ref{i1i2} summarizes results of the Welch's $t$-tests. We observe the same trend as we did with the exponential decay rates: in both the sonata dataset ($t$-test, $t$=6.24, $df$ = 57.03, and $p$-value = $5.85\times10^{-8}$) and the quartet dataset ($t$=3.51, $df$ = 49.09, and $p$-value = $9.61\times10^{-4}$), Mozart's pieces show less predictability even when comparing $1$-note and $2$-note lags. Moreover, these differences are observed even for short time lags, indicating that even lower-order Markov models may be informative of discriminating features of a composer. This observation motivates a more direct analysis of first- and second-order Markov models in the subsequent section~\ref{markov_models}. In Supplementary Table~\ref{pair_I1I2}
, we report the full set of pairwise comparisons, which confirm consistent differences between Mozart and others, while also showing variation in how Beethoven, Haydn, and Schubert compare with one another.

\begin{table}[h!]
\centering
\caption{\textbf{Group comparisons for I(1)/I(2).} For each dataset (sonatas and quartets), we report the mean and standard deviation in the ratio, along with the number of pieces ($N$) examined for Mozart and for the combined group of other composers. We also report results from a Welch's $t$-test to evaluate whether the mean I(1)/I(2) ratio for Mozart differs significantly from the other composers.}
\label{i1i2}
\begin{tabular}{llcccc}
\toprule
Dataset & Group & Mean & Std. Dev. & N & $t$-test (Mozart vs Others) \\
\midrule
Sonata & Mozart & 1.8283 & 0.3915 & 45 & \multirow{2}{*}{$t = 6.24,\, p = 5.85\times10^{-8},\, df = 57.03$} \\
       & Others (H+B+S) & 1.4352 & 0.3679 & 278 & \\
\midrule
Quartet & Mozart & 1.7116 & 0.2885 & 34 & \multirow{2}{*}{$t = 3.51,\, p = 9.61\times10^{-4},\, df = 49.09$} \\
       & Others (H+B) & 1.5163 & 0.3309 & 194 & \\
\bottomrule
\end{tabular}
\end{table}

\subsection{Markov models} \label{markov_models}

To complement our study of the time-delayed mutual information in Sec.~\ref{sec:MI_results}, we now examine dependencies between notes in Markov models of musical sequences. While mutual information provides a summary statistic of how dependencies vary with lag, a Markov model is a generative framework that can, in principle, be used to produce new musical sequences. Such models contain richer information about musical structure than mutual information alone. Specifically, we fit first- and second-order Markov models to each piece in our dataset. We then evaluate the relative performance of these models for each piece using a log-likelihood ratio (LLR) test and the Bayesian Information Criteria (BIC) test. We seek to quantify the degree to which accounting for two prior notes provides additional predictive power than a single prior note alone. Accordingly, we calculate $\Delta$BIC, which we define as the difference in BIC values between a first- and second-order Markov model, as defined in Eq.~\eqref{eq:delta_BIC}.

\begin{figure}[h!]
    \centering
    \includegraphics[width=0.8\linewidth]{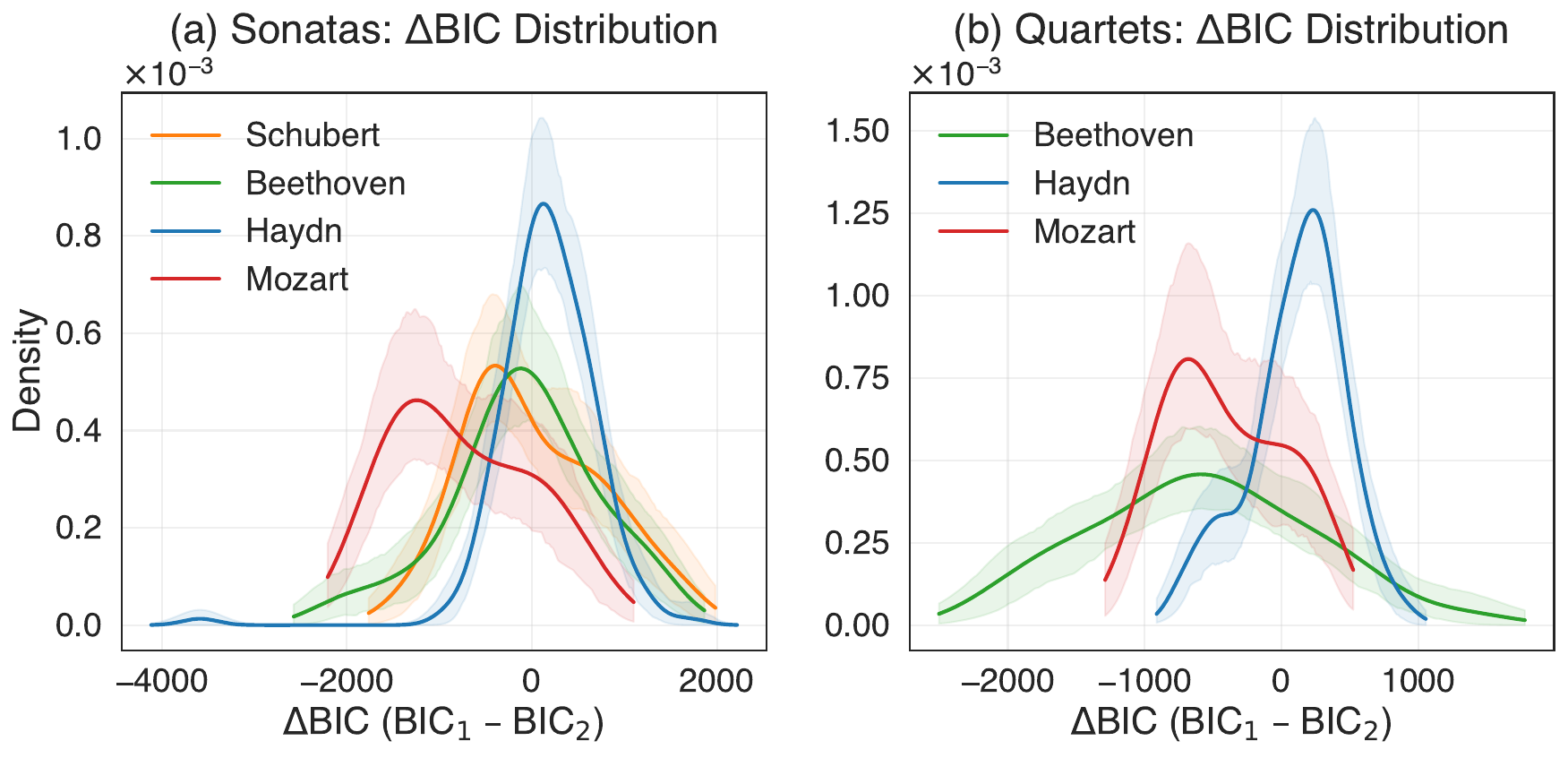}
    \caption{\textbf{Density plots of the distributions of $\Delta$BIC values for Markov-1 versus Markov-2 fits.} (a) BIC values for the sonatas of four composers. (b) BIC values for the quartets of three different composers.}
    \label{fig:delta_BIC_fig}
\end{figure}

The distributions of $\Delta$ BIC for each composer are shown in Fig.~\ref{fig:delta_BIC_fig}. The summary statistics for the LLR distributions are reported in Table~\ref{sum_llr}, and for $\Delta$BIC in Table~\ref{sum_delta_BIC}. Because the LLR increases with sequence length, it can be biased by piece size; this effect is evident in our dataset, where LLR is correlated with piece length but $\Delta$BIC is not (see Supplementary Fig.~\ref{LL_len}). For this reason, we primarily report comparisons based upon the BIC results in the main text. The density plots and the summary statistics of the LLRs are reported in the Supplement Fig.~\ref{fig:LLR_densityplot} and Table~\ref{sum_llr}.

For sonatas, we observe that Mozart's pieces are shifted towards negative BIC values compared to other composers, indicating better fits with first-order models. This trend is also reflected in the summary counts: first-order models are preferred by the BIC test for $78\%$ of Mozart sonatas compared to 56\% of Beethoven, 32\% of Haydn, and 55\% of Schubert sonatas (see Table ~\ref{table:BIC_LLR_summary}). The LLR produced a similar pattern, preferring first-order models for approximately 84\% of Mozart sonatas, compared to 46\%, 55\%, and 29\% of Beethoven, Haydn, and Schubert sonatas. For quartets, the differences are less pronounced. By BIC, first-order models were selected for $74\%$ of Mozart quartets, a proportion comparable to Beethoven ($74\%$) and higher than Haydn ($35\%$). 

We perform a Welch's $t-$test to test for differences in the mean $\Delta$BIC values between Mozart and other composers, and find significant differences for both the sonatas ($t = -4.65,\, p < 2\times10^{-5},\, df = 88.10$) and the quartets ($t = -3.01,\, p = 0.004,\, df = 61.75$), as reported in Table~\ref{bic}. However, as shown in Fig.~\ref{fig:delta_BIC_fig}(b), the differences across composers for the string quartets appear more nuanced. In Supplementary Table~\ref{pair_delta_bic}, we report results from a Welch's $t$-test to test for differences between each pair of composers. We observe that while Mozart differs from all other composers in the Sonatas, the differences between Mozart and Beethoven's string quartets are not that significant. Together, these findings support the view that note transitions in Mozart’s compositions are more strongly captured by first-order dependencies, whereas second-order models provide better fits for Haydn and Schubert, with Beethoven showing an intermediate pattern that varies by form.

\begin{table}[h!]
\centering
\caption{\textbf{Group comparisons of the LLR.} For each dataset (sonatas and quartets), we report the mean and standard deviation of the LLR (Markov-1 versus Markov-2), along with the number of pieces ($N$) examined for Mozart and for the combined group of other composers. Welch's $t$-tests evaluate whether the mean LLR value for Mozart differs significantly from the other composers.}
\label{llr}
\begin{tabular}{llcccc}
\toprule
Dataset & Group & Mean & Std. Dev. & N & $t$-test (Mozart vs Others) \\
\midrule
Sonata & Mozart & 1263.0777 & 610.9288 & 45 & \multirow{2}{*}{$t = 
-4.65,\, p < 2\times10^{-5},\, df = 88.10$} \\
       & Others (H+B+S) & 1776.4308 & 1012.4792 & 278 & \\
\midrule
Quartet & Mozart & 1008.3823 & 479.3164 & 34 & \multirow{2}{*}{$t = 
-3.76,\, p < 4\times10^{-4},\, df = 69.82$} \\
       & Others (H+B) & 1390.6869 & 805.8790 & 194 & \\
\bottomrule
\end{tabular}
\end{table}

\begin{table}[h!]
\centering
\caption{\textbf{Group comparisons for $\Delta$BIC.} For each dataset (sonatas and quartets), we report the mean and standard deviation of the computed $\Delta$BIC values, along with the number of pieces ($N$) examined for Mozart and for the combined group of other composers. Welch's $t$-tests evaluate whether the $\Delta$BIC value for Mozart differs significantly from the other composers.}
\label{bic}
\begin{tabular}{llcccc}
\toprule
Dataset & Group & Mean & Std. Dev. & N & $t$-test (Mozart vs Others) \\
\midrule
Sonata & Mozart & -751.7989 & 755.5530 & 45 & \multirow{2}{*}{$t = 
-6.94,\, p < 10^{-8},\, df = 54.82$} \\
       & Others (H+B+S) & 83.4394 & 649.5142 & 278 & \\
\midrule
Quartet & Mozart & -405.4947 & 431.0613 & 34 & \multirow{2}{*}{$t = -3.01,\, p = 0.004,\, df = 61.75$} \\
       & Others (H+B) & -140.0453 & 646.9220 & 194 & \\
\bottomrule
\end{tabular}
\end{table}

\begin{table}[htbp]
\centering
\begin{tabular}{llcccc}
\toprule
\textbf{Composer} & \textbf{Dataset} & \textbf{Criterion} & \textbf{Markov-1} & \textbf{Markov-2} & \textbf{Total} \\
\midrule
Mozart    & Sonata  & BIC & 35 (77.8\%) & 10 (22.2\%) & 45 \\
          &         & LLR & 38 (84.4\%) & 7 (15.6\%)  &  \\
          & Quartet & BIC & 25 (73.5\%) & 9 (26.5\%)  & 34 \\
          &         & LLR & 30 (88.2\%) & 4 (11.8\%)  &  \\
\midrule
Beethoven & Sonata  & BIC & 34 (55.7\%) & 27 (44.3\%) & 61 \\
          &         & LLR & 28 (45.9\%) & 33 (54.1\%) &  \\
          & Quartet & BIC & 52 (74.3\%) & 18 (25.7\%) & 70 \\
          &         & LLR & 31 (44.3\%) & 39 (55.7\%) &  \\
\midrule
Haydn     & Sonata  & BIC & 52 (32.7\%) & 107 (67.3\%) & 159 \\
          &         & LLR & 88 (55.3\%) & 71 (44.7\%)  &  \\
          & Quartet & BIC & 43 (34.7\%) & 81 (65.3\%)  & 124 \\
          &         & LLR & 101 (81.5\%) & 23 (18.5\%) &  \\
\midrule
Schubert  & Sonata  & BIC & 32 (55.2\%) & 26 (44.8\%) & 58 \\
          &         & LLR & 17 (29.3\%) & 41 (70.7\%) &  \\

\bottomrule
\end{tabular}
\caption{\textbf{Counts and proportions of sonata and quartet movements best fit by first- or second-order Markov models.} Here we evaluate fit according to the Bayesian Information Criterion (BIC) and Likelihood Ratio Test (LRT), for each composer.}
\label{table:BIC_LLR_summary}
\end{table}

\subsection{MTD models}

To complement our tests based on fitting first- and second-order Markov models, we now fit mixture transition distribution (MTD) models to the same corpus of sonatas and quartets. In this framework, the conditional probability of the next note is expressed as a weighted mixture of contributions from multiple preceding notes. The weights $\beta_{k}$ indicate how much predictive power is assigned to the note at lag k. Here, we focus on second-order (MTD-2) and third-order (MTD-3) fits, which allow us to examine the relative influence of lag-1 compared with lags of two or three notes back. After fitting an MTD to each piece in our dataset, we assess the extent of statistical dependencies in note sequences by inspecting the lag-1 weight $(\beta_1)$, which reflects the contribution of the immediately preceding note. In our analysis, we compare the distributions of $\beta_1$ across composers for MTD-2 and MTD-3 fits. In addition, we highlight individual movements in our sonata dataset where longer-time weights ($\beta_2$ or $\beta_3$) are especially pronounced, as an illustrative example of extended dependencies.

\begin{figure}
    \centering
    \includegraphics[width=0.8\linewidth]{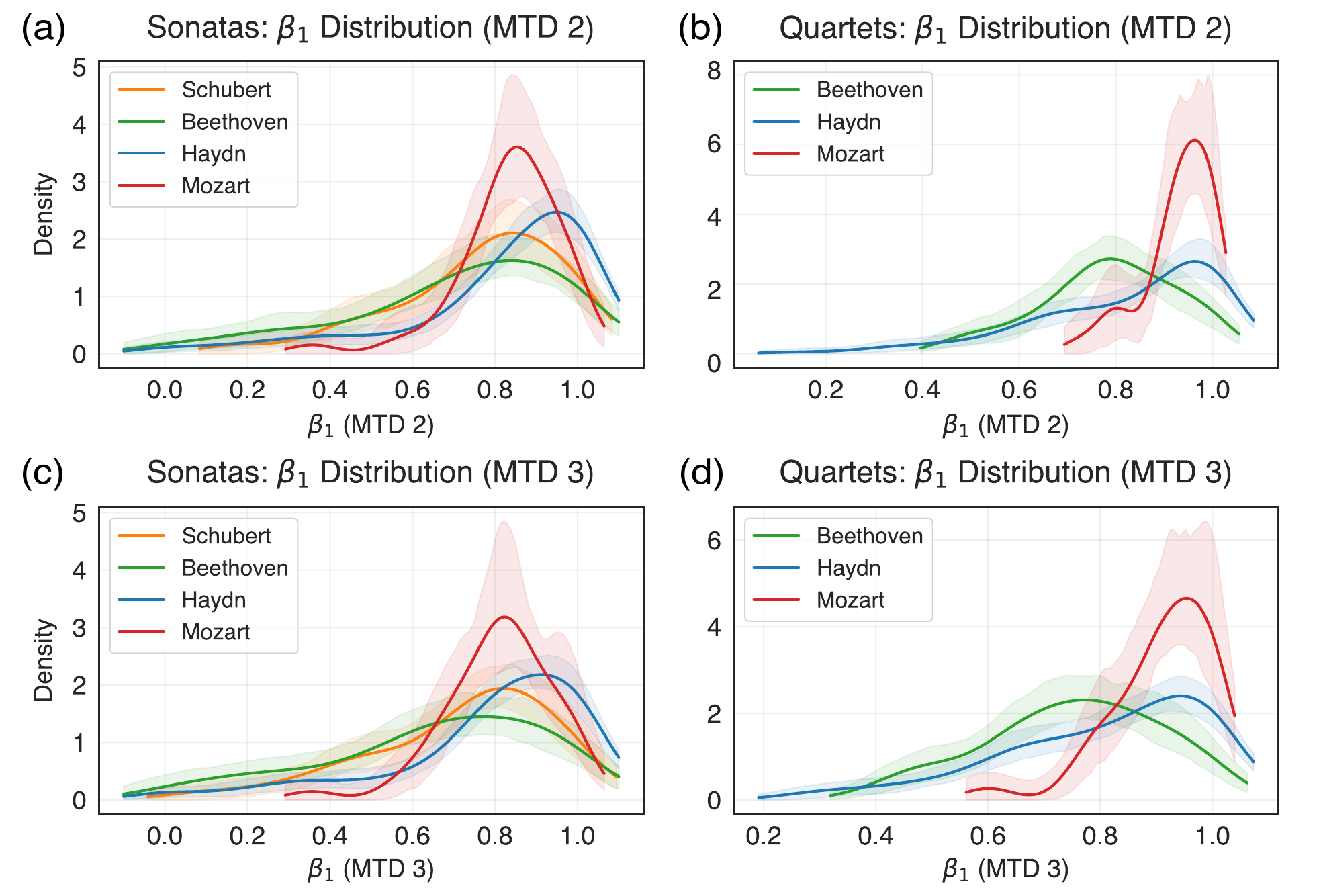}
    \caption{\textbf{Density plots of the distributions of $\beta_1$ parameters for MTD models.} (a) Density plots of $\beta_1$ parameters of second-order MTDs in the sonata dataset. (b) Density plots of $\beta_1$ parameters of second-order MTDs in the quartet dataset. (c) Density plots of $\beta_1$ parameters of third-order MTDs in the sonata dataset. (d) Density plots of $\beta_1$ parameters of third-order MTDs in the quartet dataset.}
    \label{MTD_plots}
\end{figure}

In Fig.~\ref{MTD_plots}, we show the distributions of lag-1 weights ($\beta_1$) obtained from MTD-2 (panels (a)--(b)) and MTD-3 (panels (c)--(d)) for the sonatas and quartets. The summary statistics of the obtained $\beta_{1}$ distributions are reported in Tables~\ref{sum_beta1_mtd2} and ~\ref{sum_beta1_mtd3}, and group comparisons between Mozart and the other composers are given in Tables~\ref{beta1mtd2} and ~\ref{beta1mtd3}. For the quartets, the lag-1 weights of Mozart are shifted towards high values, with a mean around $0.91$ compared to $0.80$ for Haydn and $0.75$ for Beethoven. The reported Welch's $t-$tests confirm that these differences are significant. For the sonatas, Mozart also tends to have higher $\beta_1$ values, but the differences related to Haydn and Schubert are less pronounced, as seen from the pairwise corrections reported in Tables~\ref{pair_beta1_MTD2} and ~\ref{pair_beta1_MTD3}. Overall, these results are broadly consistent with our earlier findings, but the contrasts among composers vary across metrics, reflecting that although all of these measures capture aspects of higher-order dependencies, they emphasize different facets of structure.

\begin{table}[h!]
\centering
\caption{\textbf{Group comparisons for $\beta_1$ (MTD-2).} For each dataset (sonatas and quartets), we report the mean and standard deviation of the computed $\beta_1$ (MTD-2) values, along with the number of pieces ($N$) examined for Mozart and for the combined group of other composers. Welch's $t$-tests evaluate whether the mean $\beta_1$ (MTD-2) value for Mozart differs significantly from the other composers.}
\label{beta1mtd2}
\begin{tabular}{llcccc}
\toprule
Dataset & Group & Mean & Std. Dev. & N & $t$-test (Mozart vs Others) \\
\midrule
Sonata & Mozart & 0.8359 & 0.1214 & 45 & \multirow{2}{*}{$t = 2.54,\, p = 0.012,\, df = 109.43$} \\
       & Others (H+B+S) & 0.7766 & 0.2412 & 279 & \\
\midrule
Quartet & Mozart & 0.9314 & 0.0713 & 34 & \multirow{2}{*}{$t = 6.87,\, 
p < 10^{-9},\, df = 113.80$} \\
       & Others (H+B) & 0.8103 & 0.1735 & 194 & \\
\bottomrule
\end{tabular}
\end{table}

\begin{table}[h!]
\centering
\caption{\textbf{Group comparisons for $\beta_1$ (MTD-3).} For each dataset (sonatas and quartets), we report the mean and standard deviation of the computed $\beta_1$ (MTD-3) values, along with the number of pieces ($N$) examined for Mozart and for the combined group of other composers. Welch's $t$-tests evaluate whether the mean $\beta_1$ (MTD-3) value for Mozart differs significantly from the other composers.}
\label{beta1mtd3}
\begin{tabular}{llcccc}
\toprule
Dataset & Group & Mean & Std. Dev. & N & $t$-test (Mozart vs Others) \\
\midrule
Sonata & Mozart & 0.8087 & 0.1275 & 45 & \multirow{2}{*}{$t = 2.90,\, p = 0.005,\, df = 107.89$} \\
       & Others (H+B+S) & 0.7379 & 0.2507 & 279 & \\
\midrule
Quartet & Mozart & 0.9114 & 0.0879 & 34 & \multirow{2}{*}{$t = 6.39,\, 
p < 10^{-8},\, df = 86.04$} \\
       & Others (H+B) & 0.7847 & 0.1749 & 194 & \\
\bottomrule
\end{tabular}
\end{table}

\begin{figure}
    \centering
    \includegraphics[width=0.9\linewidth]{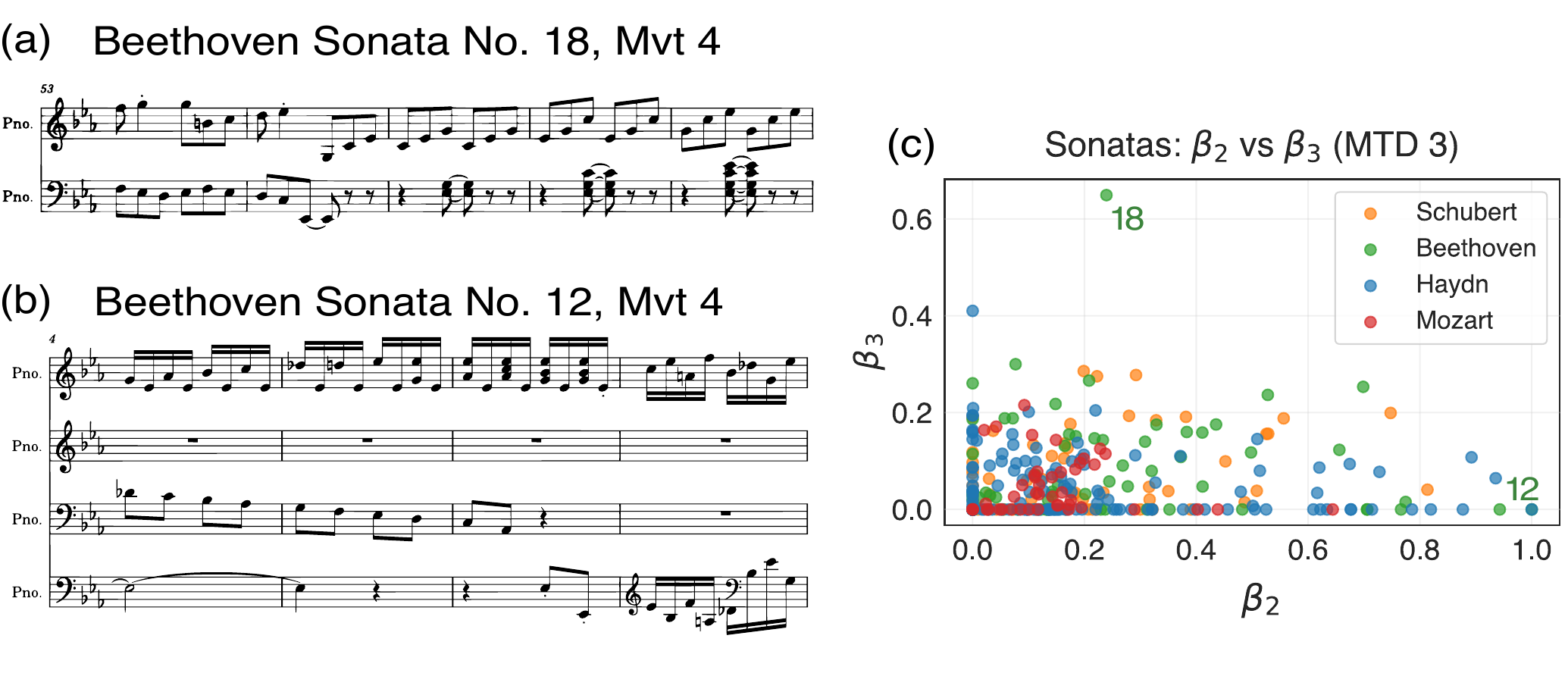}
    \caption{ \textbf{Example sonatas with especially high $\beta_2$ or $\beta_3$ values.} (a) The fourth movement of Beethoven's Sonata No.18. (b) The fourth movement of Beethoven's Sonata No. 12. (c) Scatter plot of the lag weights of third-order MTDs in the sonata dateset, with the two movements of interest pointed out.}
    \label{MTD_examples}
\end{figure}

We now highlight some movements in our dataset with especially high $\beta_2$ or $\beta_3$ values. For example, the fourth movement of Beethoven's Sonata No.12, nicknamed ``Funeral March", has $\beta_2 = 0.94$. The structure of the piece explains why: in this movement Beethoven frequently employs an alternating pattern in which a single note is repeated in-between non-repeated notes (see Fig \ref{MTD_examples}b). Likewise, the fourth movement of Beethoven's Sonata No.18, nicknamed ``The Hunt", has $\beta_3 = 0.65$ (higher than any other sonata movement). Once again, examining the music explains the reason for this result: the movement---which is in 6/8 time specifically to accommodate the following feature---is built on a near-constant backbone of repeated three-note phrases (see Fig \ref{MTD_examples}a). These examples illustrate that MTD models, and the weights associated with different orders, capture nontrivial melodic structures.

\section{Conclusion and Discussion}

In this study, we used information theory and probabilistic modeling to quantify statistical dependencies in piano sonatas and string quartets composed by Mozart, Beethoven, Haydn, and Schubert. Our dataset comprised $316$ piano sonatas and $289$ string quartets, from which we examined melody lines by extracting sequences of top notes and mapping them onto a single octave. We then assessed statistical structure using three distinct measures: first- versus second-order Markov model fits, time-delayed mutual information, and mixture transition distribution (MTD) models. Across these approaches, we observed that Mozart’s compositions generally exhibit a stronger emphasis on immediate note-to-note dependencies than those of the other composers. At the same time, there are nuances across compositional forms and metrics: in the quartets, some of Beethoven's compositions show a similar evidence of stronger local dependencies, while in the sonatas the MTD models indicate comparable short-range structure for Haydn and Mozart.

Through time-delayed mutual information analysis, we found that Mozart's compositions demonstrate relatively faster decay rates of predictive information across temporal lags, indicating that melodic events become independent more quickly than in works by other composers. We then fit markov chain models of order $1$ and $2$ and quantified their relative fits using the Bayesian Information Criterion. For the sonatas, second-order models provided better fits for Beethoven, Haydn, and Schubert, whereas first-order models generally provided better fits for Mozart. In the quartets, second-order models provide better fits for Haydn, but first-order models provide better fits for Mozart and some compositions by Beethoven. Finally, we also fit mixture transition distribution models of orders $2$ and $3$, and examined the weights assigned to each temporal lag. Here, the coefficients for lag 1 were consistently higher for Mozart's quartets. For the sonatas, lag 1 weights were comparable across Mozart, Haydn, and Schubert. Taken together, these analyses suggest that Mozart’s style often emphasizes highly localized statistical dependencies, with melodic events more strongly predicted by the immediately preceding note than by longer temporal contexts, with nuances that depend on the compositional form. In this sense, Mozart's works exhibits relatively shallow statistical dependencies. This quantitative characterization of Mozart's distinctive compositional fingerprint provides a foundation for future investigations of how varying extents of statistical dependencies in musical composition may influence listener perception, aesthetic appreciation, and cognitive processing of musical structure.

\subsection{Early applications of information theory and Markov chains to music}

Within just a few years after the publication of ``A Mathematical Theory of Communication,'' a book which introduced the fundamental principles of information theory---including entropy and mutual information---scholars began to explore the application of these concepts to music. Compared to other early domains of information theory application, such as telephone signal transmission \cite{shannon1948mathematical}, data compression \cite{huffman1952method}, and cryptography \cite{shannon1949communication}, music seems a more abstract and less direct fit for these quantitative methods. Indeed, the application of information theory to music was unique because it did not offer an immediate tangible utility in the sense that the other applications did, which were driven by practical engineering or security concerns. Instead, music attracted researchers out of a desire to better understand an aesthetic and cultural phenomenon through quantitative means. By 1956, Markov chains were applied in tandem with information theory \cite{meyer1956emotion, pinkerton1956}. Before then, researchers had explored the use of Markov chains to generate novel compositions \cite{hiller1959experimental, brooks1957experiments}. These early studies are of interest to us because the methods they employ are in many ways much more alike to ours than the methods commonly used in modern music cognition. Our modeling techniques and information theoretic measures are chosen to be interpretable, in contrast to black-box approaches such as large language models and hidden Markov models.

\subsection{The inverted‑U hypothesis for music enjoyment}

In psychology, the inverted-U hypothesis refers to the idea that enjoyment of a stimulus is highest at moderate levels of arousal or complexity—too little leads to boredom, too much leads to confusion or stress \cite{berlyne1971aesthetics}. Empirical studies in visual art, consumer product design, humor, and narrative media are consistent with this hypothesis, finding that moderate complexity tends to elicit the highest aesthetic ratings, while both extreme simplicity and excessive complexity reduce preference \cite{north1995subjective, gucluturk2016liking}. In music cognition, the concept of the inverted-U relationship between complexity and enjoyment has been developed in terms of expectation. Meyer’s influential theory proposed that musical emotions arise from the confirmation or violation of learned expectations based on prior exposure to musical structure \cite{meyer1956emotion}. More recent work has translated this idea into formal probabilistic and information-theoretic models, where listener expectations are modeled using statistical learning, and quantities such as entropy and surprisal are used to quantify uncertainty and unexpectedness in musical sequences \cite{pearce2012auditory}.

Experimental studies have provided converging evidence that musical pleasure follows a non-monotonic function of uncertainty and surprise. Gold \emph{et al.} (2019) found that musical excerpts eliciting high enjoyment tend to fall into regions of intermediate surprisal in a low-uncertainty context, or into regions of low surprisal in a high-uncertainty context—consistent with the idea that musical pleasure results from a balance between predictability and novelty \cite{gold2019uncertainty}. Neuroimaging research supports this interpretation: interactions between predictive variables and subjective pleasure are reflected in activity within the auditory cortex, amygdala, hippocampus, and nucleus accumbens \cite{zatorre2013perception, salimpoor2011dopamine}. In the context of statistical dependencies, which is another form of predictability, one may expect the most popular composers to strike a balance of some sort. It is possible that this balance could change over time as musical tastes (and predictions) evolve alongside culture, and differ in more local geographic contexts. In light of this possibility, it could be of interest in future to consider temporal and cultural variation in statistical predictability of musical compositions, and the degree to which it might be expected that Mozart, given his time and place, tended towards notes that were predicted by more local note-to-note transitions.


\subsection{Methodological considerations}

Several methodological considerations are pertinent to the nature and generalizability of our findings. Our analysis currently relies on extracting top notes from polyphonic scores. This approach made our analysis tractable, but introduces limitations. The method may incorrectly identify ornamentation, accompaniment, and non-melodic high notes as parts of the melody. Moreover, by reducing each piece to a monophonic line mapped into a single octave, we discard important aspects of musical structure such as rhythm, tempo, harmony, timbre, volume, duration, and pitch range. We particularly note that mapping the notes to a single octave can make large jumps considered equal to small jumps, despite the fact that large jumps are often far more sparse and surprising.

Another consideration relates to the simplifications that each statistical method entails. Standard Markov models treat past notes as fully determinative of the next, while MTD models assume additive influences of prior lags. These assumptions may not always hold and could miss musical features. There are also certain common musical features -- such an inverted motif -- that rely on abstract structural transformations that a Markov model would not predict. Such musical features are central to compositional style and listener experience, but they fall outside the representational scope of the models we employ here.

We also note that all three of our measures of statistical dependency (mutual information, Markov fits, and MTD weights) are computed on entire pieces. This procedure ignores within-piece heterogeneity and contextual rules: changes in key, form, or theme that shape how expectations unfold over time. In tonal music, for example, certain transitions are strongly constrained by harmonic rules, modulations, or cadential patterns, while others are permitted only in specific sections of a piece. A human listener, for example, would be able to tell that certain transitions are unlikely in the context of a given key, and would be able to discriminate between separate themes.

\subsection{Further Directions: Dynamic approaches to modeling memory and surprise}

Our work highlights differences in temporal dependencies across composers and forms. An interesting direction for future research is to develop models of perception that account for how expectations vary dynamically, rather than being captured by a fixed-order Markov model. Such approaches could better reflect the possibility that listeners rely on different temporal horizons depending on context, style, or experience.

To capture such a possibility, it would be of interest to consider an approach that blends multiple sources of statistical expectation through a mixture of Markov-\(k\) models over distinct time windows. For example, this mixture could be built to predict a listener's evolving perception at time \( t \) by combining three separate models: a stylistic model, a piece model, and a context model. The stylistic model, based on a large corpus of music, could reflect general expectations shaped by a listener’s exposure to a composer or style. The piece model could track patterns that have emerged over the course of the particular piece the listener is currently hearing. And the context model could focus only on the most recent notes, thereby capturing immediate patterns or motifs. At each moment in the piece, the listener’s expectations about what note is likely to come next could be modeled as a weighted combination of these three sources. The exact balance among them can shift depending on the listener’s experience or the structure of the music. This approach is more nuanced than a simple Markov model trained on a single piece, and may better reflect the layered and evolving nature of musical expectation.

This type of framework could then be connected to perceptual and emotional responses, and it would be of interest to consider two commonly used informational quantities in models of musical cognition: uncertainty and surprise. Uncertainty, or entropy, represents how unpredictable the next note is from the listener’s perspective. If many different notes seem equally likely, then the uncertainty is high; if one note stands out as highly probable, then uncertainty is low. Surprise, or information content, reflects how unexpected the actual next note turns out to be. A note that was judged likely will produce little surprise; a note that seemed improbable will produce more surprise. According to the inverted-U hypothesis in music psychology, enjoyment is not driven solely by uncertainty or surprise in isolation, but by their interaction. Enjoyment tends to be highest when expectations are confidently formed and then cleverly violated, or when uncertainty is high and the music provides an elegant or clarifying resolution. By contrast, when both uncertainty and surprise are low, the music may feel predictable or dull. And when both are high, the result may feel chaotic or overwhelming.

A separate but potentially also fruitful approach could be to employ variable-length Markov models (VLMMs), which generalize fixed-order models by allowing the memory depth to adapt based on the statistical regularities of the input. That is, rather than committing to a single Markov-\(k\) model, a VLMM builds a context tree where longer histories are used when supported by sufficient data, while shorter contexts are used when evidence is sparse \cite{rissanen1983vlmm,begleiter2004prediction,spiliopoulou2011dirichlet}. Importantly, VLMMs produce an explicit, data-driven structure (a context tree or PST) that reveals which past sequences are influential, which makes them more interpretable than black‑box methods such as HMMs or neural networks. In 1995, Conklin \& Witten applied VLMMs to a musical corpus of 95 Bach chorales for the purpose of music generation, and found that VLMMs generated superior compositions in comparison to standard Markov models \cite{conklin1995multiple}, but the use of VLMMs in an interpretive capacity was left unexplored for music. Applied to music cognition, VLMMs offer a promising route for modeling how memory itself becomes context-sensitive, encoding the idea that the listener’s window of attention expands or contracts depending on the local predictability of the musical surface. Future work could replace the fixed-$k$ models with VLMMs that reflect not only how much memory is used, but when and where that memory is applied. This opens the door to a richer and more cognitively plausible understanding of how memory, expectation, and surprise interact in musical perception.

\section{Code and data availability}
The computer code required to execute and reproduce these results is available on Github:  \url{https://github.com/lchenplotkin/markov-Mozart}.

\printbibliography

\clearpage
\section{Supplementary Material}

\setcounter{equation}{0}
\setcounter{figure}{0}
\setcounter{table}{0}
\setcounter{page}{1}
\makeatletter
\renewcommand{\theequation}{S\arabic{equation}}
\renewcommand{\thefigure}{S\arabic{figure}}
\renewcommand{\thetable}{S\arabic{table}}

In this Supplementary Materials section, we provide additional figures and tables for the analyses reported in the main text. In Sec.~\ref{sec:LLR_supp}, we show the density plots of the measured LLRs in our dataset and examine the correlation between the LLR and $\Delta$BIC values of each piece with the length of the piece. In Sec.~\ref{sec:summary_all}, we report summary statistics for the distributions of all measured quantities. In Sec.~\ref{sec:pairwise_all}, we present pairwise comparisons between composers for each metric. Finally, in Sec.~\ref{sec:grouped_all}, we report comparisons of each composer against the combined set of the other composers.

\subsection{Log-likelihood ratio (LLR) plots} \label{sec:LLR_supp}

\begin{figure}[h!]
    \centering
    \includegraphics[width=0.8\linewidth]{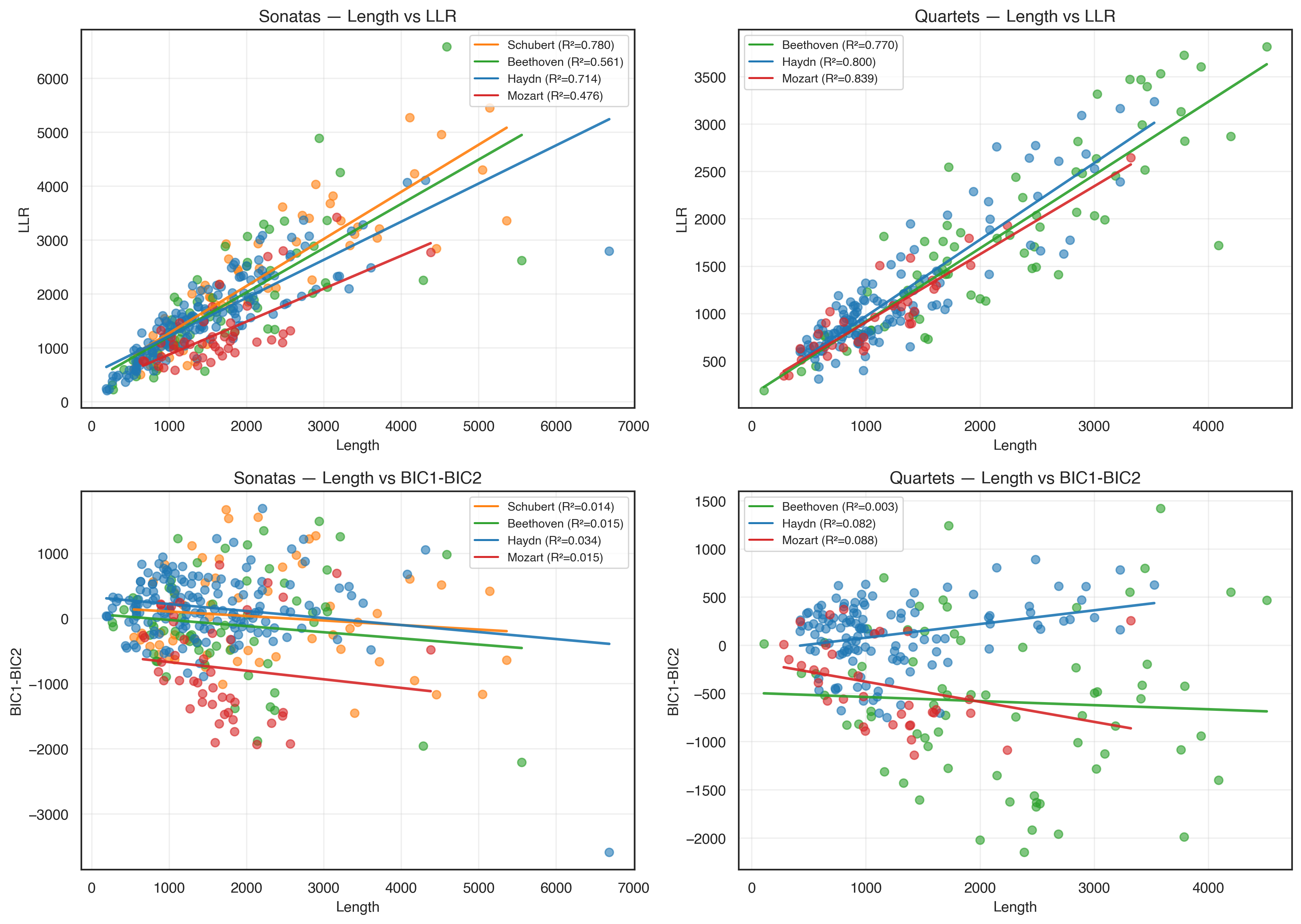}
    \caption{\textbf{Correlation between the log-likelihood ratios (LLR) and the $\Delta$BIC values with piece length.} Top row: We examine the log-likelihood ratios as a function of the length of the pieces for the sonatas (left) and quartets (right). Bottom row: We examine the $\Delta$BIC values as a function of the length of the pieces for the sonatas (left) and quartets (right). In each panel, lines of best fit are drawn separately for each composer, with associated $R^2$ values reported in the legend.}
    \label{LL_len}
\end{figure}

In Sec.~\ref{LLR_BIC_methods}, we noted that log-likelihood test results are influenced by sequence length. To address this dependence, we considered the Bayesian Information Criterion (BIC), which introduces a penalty term that adjusts for sequence length. We test this claim empirically in our dataset by examining how the LLR and $\Delta$BIC values are correlated with piece length. In Fig.~\ref{LL_len}, we show scatter plots of the LLR (top row) and $\Delta$BIC values (bottom row) for the sonatas and quartets. For each composer, we fit regression lines and report the associated $R^{2}$ values. Across both types of compositions, we observe that the LLR and piece length are correlated (with $R^2$ values as high as 0.8), whereas $\Delta$BIC and piece length are not as strongly correlated ($R^2 < .05$).

\begin{figure}[h!]
    \centering
    \includegraphics[width=0.8\linewidth]{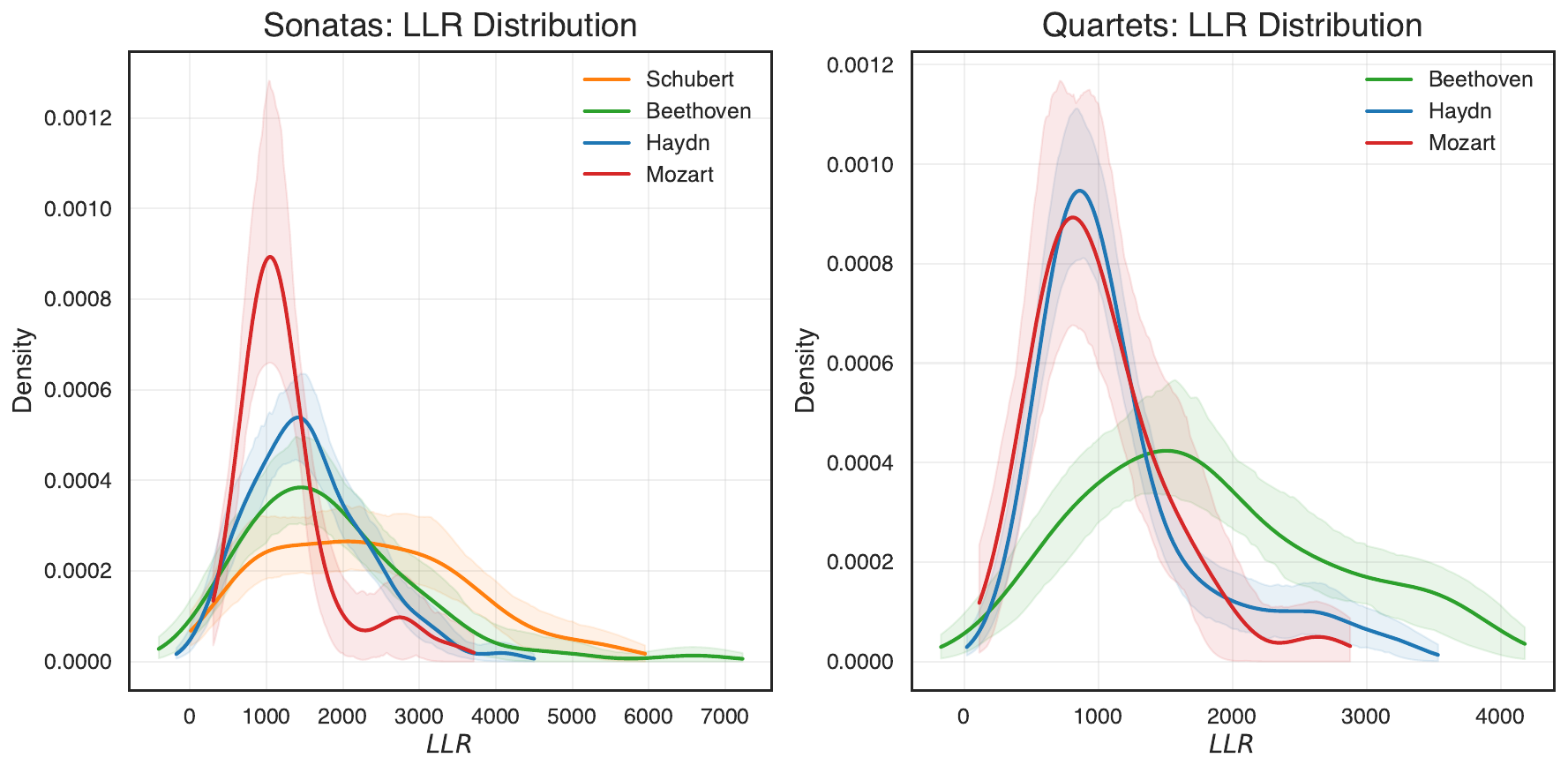}
     \caption{\textbf{Density plots of the distributions of log-likelihood values in our dataset.} (a) LLR values for the sonatas of the four composers. (b) LLR values for the quartets of the three different composers.}
    \label{fig:LLR_densityplot}
\end{figure}

For this reason, we emphasize BIC-based tests in the main text. In addition to the density plots of $\Delta$BIC (Figure \ref{fig:delta_BIC_fig}), we also provide density plots of the raw LLR values for each composer (Fig.~\ref{fig:LLR_densityplot}), though we note that these are dependent on piece length.

\subsection{Summary statistics for all metrics}\label{sec:summary_all}

Tables \ref{sum_decay}--\ref{sum_beta3_mtd3} present summary statistics for the $8$ metrics of statistical dependency ($\lambda,  I(1)/I(2)$, LLR, $\Delta$BIC, coefficients $\beta_{1}$ from a second-order MTD model, and coefficients $\beta_{1}, \beta_{2}, \beta_{3}$ from a third-order MTD model). We note that the qualitative trends are not consistent across all metrics. 

For the sonatas, Mozart shows the strongest evidence of localized dependencies across multiple measures. The mean values are highest for the decay constant $ \lambda $, the ratio $ I(1)/I(2) $, and the lag-1 weights $ \beta_{1} $ of the second and third order MTD fits. The mean $\Delta$ BIC value is also strongly negative, indicating that first-order models are more often preferred for Mozart than for the other composers. At the same time, however, the lag-1 weights $ \beta_{1} $ for Mozart are close to Haydn’s, while the log-likelihood ratio (LLR) shows the lowest mean for Mozart.

For the quartets, Mozart again shows higher values for $\lambda$, $I(1)/I(2)$, and $ \beta_{1}$, with mean lag-1 weights exceeding $ 0.9 $ in the MTD fits. Although the $ \Delta$ BIC values are negative (consistent with a preference for first-order models), we observe an even lower mean value for quartets composed by Beethoven. 

Taken together, these results suggest that Mozart’s music is generally characterized by stronger reliance on immediate note-to-note transitions, although the exact ranking among composers varies by metric. This observed behavior motivates the pairwise and group comparisons in Sec.~\ref{sec:pairwise_all} and Sec.~\ref{sec:grouped_all}, which provide a fuller picture of how statistical dependencies differ across composers.

\begin{table}[h!]
\centering
\caption{\textbf{Summary Statistics for the measured decay constants $(\lambda s)$.} For each composer and dataset (sonatas and quartets), we report the mean, standard deviation, sample size $(N)$, and range (minimum $–$ maximum) of the fitted decay constants.}
\label{sum_decay}
\begin{tabular}{llcccc}
\toprule
Dataset & Composer & Mean & Std. Dev. & N & Min--Max \\
\midrule
Sonata & Mozart & 0.2477 & 0.1112 & 45 & 0.089--0.535 \\
 & Haydn & 0.1329 & 0.0612 & 159 & 0.018--0.349 \\
 & Beethoven & 0.1446 & 0.0688 & 61 & 0.036--0.365 \\
 & Schubert & 0.1699 & 0.0651 & 58 & 0.057--0.362 \\
\midrule
Quartet & Mozart & 0.2337 & 0.1235 & 34 & 0.059--0.550 \\
 & Haydn & 0.1591 & 0.0704 & 124 & 0.036--0.347 \\
 & Beethoven & 0.1762 & 0.0781 & 70 & -0.003--0.424 \\
\bottomrule
\end{tabular}
\end{table}

\begin{table}[h!]
\centering
\caption{\textbf{Summary Statistics of $I(1)/I(2)$.} For each composer and dataset (sonatas and quartets), we report the mean, standard deviation, sample size $(N)$, and range (minimum $–$ maximum) of the measured $I(1)/I(2)$ ratio.}
 \label{sum_ratio}
\begin{tabular}{llcccc}
\toprule
Dataset & Composer & Mean & Std. Dev. & N & Min--Max \\
\midrule
Sonata & Mozart & 1.8283 & 0.3915 & 45 & 0.893--2.644 \\
 & Haydn & 1.4706 & 0.3755 & 159 & 0.500--2.592 \\
 & Beethoven & 1.3463 & 0.3450 & 61 & 0.607--2.012 \\
 & Schubert & 1.4315 & 0.3540 & 58 & 0.581--2.374 \\
\midrule
Quartet & Mozart & 1.7116 & 0.2885 & 34 & 1.270--2.292 \\
 & Haydn & 1.5049 & 0.3208 & 124 & 0.749--2.385 \\
 & Beethoven & 1.5365 & 0.3471 & 70 & 0.862--2.462 \\
\bottomrule
\end{tabular}
\end{table}

\begin{table}[h!]
\centering
\caption{\textbf{Summary Statistics of the LLRs.} For each composer and dataset (sonatas and quartets), we report the mean, standard deviation, sample size $(N)$, and range (minimum $–$ maximum) of the measured LLR values.}
\label{sum_llr}
\begin{tabular}{llcccc}
\toprule
Dataset & Composer & Mean & Std. Dev. & N & Min--Max \\
\midrule
Sonata & Mozart & 1263.0777 & 610.9288 & 45 & 587.282--3426.997 \\
 & Haydn & 1555.4273 & 757.6133 & 159 & 210.651--4109.692 \\
 & Beethoven & 1835.7685 & 1137.8560 & 61 & 226.795--6589.016 \\
 & Schubert & 2319.8783 & 1243.1283 & 58 & 505.529--5454.924 \\
\midrule
Quartet & Mozart & 1008.3823 & 479.3164 & 34 & 342.504--2647.157 \\
 & Haydn & 1151.5987 & 623.9934 & 124 & 311.895--3239.900 \\
 & Beethoven & 1814.2146 & 910.7720 & 70 & 188.955--3817.186 \\
\bottomrule
\end{tabular}
\end{table}

\begin{table}[h!]
\centering
\caption{\textbf{Summary Statistics of the $\Delta$BIC values.} For each composer and dataset (sonatas and quartets), we report the mean, standard deviation, sample size $(N)$, and range (minimum $–$ maximum) of the measured $\Delta$BIC values.}
\label{sum_delta_BIC}
\begin{tabular}{llcccc}
\toprule
Dataset & Composer & Mean & Std. Dev. & N & Min--Max \\
\midrule
Sonata & Mozart & -751.7989 & 755.5530 & 45 & -1931.290--825.211 \\
 & Haydn & 172.3607 & 524.3473 & 159 & -3587.036--1690.671 \\
 & Beethoven & -93.1846 & 795.3588 & 61 & -2204.349--1495.396 \\
 & Schubert & 25.4321 & 738.3623 & 58 & -1451.688--1671.836 \\
\midrule
Quartet & Mozart & -405.4947 & 431.0613 & 34 & -1138.104--373.015 \\
 & Haydn & 111.1118 & 342.4448 & 124 & -749.154--890.510 \\
 & Beethoven & -584.9520 & 801.5293 & 70 & -2146.566--1421.448 \\
\bottomrule
\end{tabular}
\end{table}

\begin{table}[h!]
\centering
\caption{\textbf{Summary Statistics of the $\beta_{1}$ values from a second-order MTD model.} For each composer and dataset (sonatas and quartets), we report the mean, standard deviation, sample size $(N)$, and range (minimum $–$ maximum) of the measured $\beta_{1}$ values from a second-order MTD model.}
\label{sum_beta1_mtd2}
\begin{tabular}{llcccc}
\toprule
Dataset & Composer & Mean & Std. Dev. & N & Min--Max \\
\midrule
Sonata & Mozart & 0.8359 & 0.1214 & 45 & 0.357--1.000 \\
 & Haydn & 0.8114 & 0.2400 & 160 & 0.000--1.000 \\
 & Beethoven & 0.7011 & 0.2599 & 61 & 0.000--1.000 \\
 & Schubert & 0.7602 & 0.2007 & 58 & 0.167--1.000 \\
\midrule
Quartet & Mozart & 0.9314 & 0.0713 & 34 & 0.721--1.000 \\
 & Haydn & 0.8245 & 0.1888 & 124 & 0.146--1.000 \\
 & Beethoven & 0.7853 & 0.1391 & 70 & 0.451--1.000 \\
\bottomrule
\end{tabular}
\end{table}

\begin{table}[h!]
\centering
\caption{\textbf{Summary Statistics of the $\beta_{1}$ values from a third-order MTD model.} For each composer and dataset (sonatas and quartets), we report the mean, standard deviation, sample size $(N)$, and range (minimum $–$ maximum) of the measured $\beta_{1}$ values from a third-order MTD model.}
\label{sum_beta1_mtd3}
\begin{tabular}{llcccc}
\toprule
Dataset & Composer & Mean & Std. Dev. & N & Min--Max \\
\midrule
Sonata & Mozart & 0.8087 & 0.1275 & 45 & 0.356--1.000 \\
 & Haydn & 0.7805 & 0.2439 & 160 & 0.000--1.000 \\
 & Beethoven & 0.6481 & 0.2674 & 61 & 0.000--1.000 \\
 & Schubert & 0.7150 & 0.2213 & 58 & 0.053--1.000 \\
\midrule
Quartet & Mozart & 0.9114 & 0.0879 & 34 & 0.601--1.000 \\
 & Haydn & 0.8043 & 0.1829 & 124 & 0.264--1.000 \\
 & Beethoven & 0.7500 & 0.1538 & 70 & 0.381--1.000 \\
\bottomrule
\end{tabular}
\end{table}

\begin{table}[h!]
\centering
\caption{\textbf{Summary Statistics of the $\beta_{2}$ values from a third-order MTD model.} For each composer and dataset (sonatas and quartets), we report the mean, standard deviation, sample size $(N)$, and range (minimum $–$ maximum) of the measured $\beta_{2}$ values from a third-order MTD model.}
\label{sum_beta2_mtd3}
\begin{tabular}{llcccc}
\toprule
Dataset & Composer & Mean & Std. Dev. & N & Min--Max \\
\midrule
 & Mozart & 0.0475 & 0.0576 & 45 & 0.000--0.215 \\
 & Haydn & 0.0383 & 0.0619 & 160 & 0.000--0.410 \\
 & Beethoven & 0.0906 & 0.1122 & 61 & 0.000--0.649 \\
 & Schubert & 0.0778 & 0.0805 & 58 & 0.000--0.286 \\
\midrule
 & Mozart & 0.0293 & 0.0504 & 34 & 0.000--0.171 \\
 & Haydn & 0.0313 & 0.0491 & 124 & 0.000--0.191 \\
 & Beethoven & 0.0654 & 0.0652 & 70 & 0.000--0.297 \\
\bottomrule
\end{tabular}
\end{table}

\begin{table}[h!]
\centering
\caption{\textbf{Summary Statistics of the $\beta_{3}$ values from a third-order MTD model.} For each composer and dataset (sonatas and quartets), we report the mean, standard deviation, sample size $(N)$, and range (minimum $–$ maximum) of the measured $\beta_{3}$ values from a third-order MTD model.}
\label{sum_beta3_mtd3}
\begin{tabular}{llcccc}
\toprule
Dataset & Composer & Mean & Std. Dev. & N & Min--Max \\
\midrule
 & Mozart & 0.1437 & 0.1201 & 45 & 0.000--0.644 \\
 & Haydn & 0.1812 & 0.2408 & 160 & 0.000--1.000 \\
 & Beethoven & 0.2613 & 0.2506 & 61 & 0.000--1.000 \\
 & Schubert & 0.2072 & 0.1904 & 58 & 0.000--0.813 \\
\midrule
 & Mozart & 0.0593 & 0.0691 & 34 & 0.000--0.228 \\
 & Haydn & 0.1645 & 0.1777 & 124 & 0.000--0.736 \\
 & Beethoven & 0.1846 & 0.1293 & 70 & 0.000--0.525 \\
\bottomrule
\end{tabular}
\end{table}

\clearpage

\subsection{Pairwise comparisons across composers} \label{sec:pairwise_all}

In Tables ~\ref{pair_lambda}--~\ref{pair_beta3_mtd3}, we present pairwise comparisons between each composer for all the metrics considered by performing a Welch's $t$-test. Since we make multiple comparisons for each metric, we use a Bonferroni adjusted threshold of $p < 0.05/9$ to test for significance.

For the sonatas, Mozart differs significantly from Beethoven and Schubert across most metrics. Relative to both composers, Mozart shows higher decay constants ($\lambda$), higher ratios ($I(1)/I(2)$), and more negative $\Delta BIC$, with additional differences against Beethoven in the lag-1 weights ($\beta_{1}$ from MTD-2 and MTD-3), LLR, and $\beta_{3}$. Comparisons with Haydn are more limited: significant differences appear only for $\lambda$, $I(1)/I(2)$, and $\Delta BIC$, while no significant contrasts are observed for LLR, $\beta_{1}$, $\beta_{2}$, or $\beta_{3}$.

For the quartets, Mozart differs significantly from Haydn across nearly all metrics, including $\lambda$, $I(1)/I(2)$, $\Delta BIC$, and both $\beta_{1}$ and $\beta_{3}$, consistently indicating stronger reliance on local structure. In contrast, comparisons between Mozart and Beethoven are mixed: significant differences appear for $\beta_{1}$ (higher for Mozart) and for $\beta_{3}$ and LLR (both lower for Mozart), but no significant differences are observed for $\lambda$, $I(1)/I(2)$, or $\Delta BIC$.

These discrepancies show that while each metric captures aspects of statistical dependency, they do emphasize different features, which can lead to qualitatively different trends across composers. Nevertheless, the general trend that Mozart differs most consistently from the other composers—particularly in the sonatas and in contrasts with Haydn for the quartets—is consistent across metrics.

\begin{table}[h!]
\centering
\caption{\textbf{Pairwise comparisons of the mean decay constants ($\lambda$) across composers.} Welch’s $t-$tests comparing composers in the sonata and quartet datasets. We report the $t$-statistics, degrees of freedom (df), and $p$-values for each test. The sign of the $t$-statistic indicates the direction: positive means the first-listed composer has the higher mean value. Boldface indicates significance at the Bonferroni-adjusted threshold ($p < 0.05/9$).}
\label{pair_lambda}
\begin{tabular}{llllcc}
\toprule
Dataset & Composer 1 & Composer 2 & $t$-statistic & df & $p$-value \\
\midrule
Sonata & Mozart & Haydn & \textbf{6.577} & \textbf{51.6} & $\mathbf{2.41\times10^{-8}}$ \\
 & Mozart & Beethoven & \textbf{5.434} & \textbf{68.2} & $\mathbf{7.99\times10^{-7}}$ \\
 & Mozart & Schubert & \textbf{4.126} & \textbf{66.8} & $\mathbf{1.05\times10^{-4}}$ \\
 & Haydn & Beethoven & -1.157 & 98.0 & 0.250 \\
 & Haydn & Schubert & \textbf{-3.740} & \textbf{95.6} & $\mathbf{3.14\times10^{-4}}$ \\
 & Beethoven & Schubert & -2.044 & 117.0 & 0.043 \\
\midrule
Quartet & Mozart & Haydn & \textbf{3.328} & \textbf{38.9} & \textbf{0.002} \\
 & Mozart & Beethoven & 2.450 & 46.0 & 0.018 \\
 & Haydn & Beethoven & -1.508 & 131.0 & 0.134 \\
\bottomrule
\end{tabular}
\end{table}

\begin{table}[h!]
\centering
\caption{\textbf{Pairwise comparisons of the mean $I(1)/I(2)$ values across composers.} Welch’s $t-$tests comparing composers in the sonata and quartet datasets. We report the $t$-statistics, degrees of freedom (df), and $p$-values for each test. The sign of the $t$-statistic indicates the direction: positive means the first-listed composer has the higher mean value. Boldface indicates significance at the Bonferroni-adjusted threshold ($p < 0.05/9$).}
\label{pair_I1I2}
\begin{tabular}{llllcc}
\toprule
Dataset & Composer 1 & Composer 2 & $t$-statistic & df & $p$-value \\
\midrule
Sonata & Mozart & Haydn & \textbf{5.407} & \textbf{68.2} & $\mathbf{8.89\times10^{-7}}$ \\
 & Mozart & Beethoven & \textbf{6.519} & \textbf{87.6} & $\mathbf{4.37\times10^{-9}}$ \\
 & Mozart & Schubert & \textbf{5.264} & \textbf{89.5} & $\mathbf{9.59\times10^{-7}}$ \\
 & Haydn & Beethoven & 2.319 & 117.1 & 0.022 \\
 & Haydn & Schubert & 0.704 & 106.3 & 0.483 \\
 & Beethoven & Schubert & -1.318 & 116.3 & 0.190 \\
\midrule
Quartet & Mozart & Haydn & \textbf{3.566} & \textbf{56.9} & $\mathbf{7.43\times10^{-4}}$ \\
 & Mozart & Beethoven & 2.680 & 76.9 & 0.009 \\
 & Haydn & Beethoven & -0.622 & 133.8 & 0.535 \\
\bottomrule
\end{tabular}
\end{table}

\begin{table}[h!]
\centering
\caption{\textbf{Pairwise comparisons of the mean LLR values across composers.} Welch’s $t-$tests comparing composers in the sonata and quartet datasets. We report the $t$-statistics, degrees of freedom (df), and $p$-values for each test. The sign of the $t$-statistic indicates the direction: positive means the first-listed composer has the higher mean value. Boldface indicates significance at the Bonferroni-adjusted threshold ($p < 0.05/9$).}
\label{pair_llr}
\begin{tabular}{llllcc}
\toprule
Dataset & Composer 1 & Composer 2 & $t$-statistic & df & $p$-value \\
\midrule
Sonata & Mozart & Haydn & -2.656 & 85.4 & 0.009 \\
 & Mozart & Beethoven & \textbf{-3.303} & \textbf{96.2} & \textbf{0.001} \\
 & Mozart & Schubert & \textbf{-5.601} & \textbf{87.2} & $\mathbf{2.44\times10^{-7}}$\\
 & Haydn & Beethoven & -1.766 & 81.0 & 0.081 \\
 & Haydn & Schubert & \textbf{-4.360} & \textbf{72.8} & $\mathbf{4.21\times10^{-5}}$ \\
 & Beethoven & Schubert & -2.194 & 114.8 & 0.030 \\
\midrule
Quartet & Mozart & Haydn & -1.423 & 66.2 & 0.159 \\
 & Mozart & Beethoven & \textbf{-5.849} & \textbf{101.1} & $\mathbf{6.13\times10^{-8}}$ \\
 & Haydn & Beethoven & \textbf{-5.377} & \textbf{106.0} & $\mathbf{4.55\times10^{-7}}$ \\
\bottomrule
\end{tabular}
\end{table}

\begin{table}[h!]
\centering
\caption{\textbf{Pairwise comparisons of the mean $\Delta$BIC values across composers.} Welch’s $t-$tests comparing composers in the sonata and quartet datasets. We report the $t$-statistics, degrees of freedom (df), and $p$-values for each test. The sign of the $t$-statistic indicates the direction: positive means the first-listed composer has the higher mean value. Boldface indicates significance at the Bonferroni-adjusted threshold ($p < 0.05/9$).}
\label{pair_delta_bic}
\begin{tabular}{llllcc}
\toprule
Dataset & Composer 1 & Composer 2 & $t$-statistic & df & $p$-value \\
\midrule
Sonata & Mozart & Haydn & \textbf{-7.619} & \textbf{56.3} & $\mathbf{3.18\times10^{-10}}$ \\
 & Mozart & Beethoven & \textbf{-4.295} & \textbf{97.4} & $\mathbf{4.14\times10^{-5}}$ \\
 & Mozart & Schubert & \textbf{-5.177} & \textbf{93.5} & $\mathbf{1.29\times10^{-6}}$ \\
 & Haydn & Beethoven & 2.396 & 80.6 & 0.019 \\
 & Haydn & Schubert & 1.382 & 78.7 & 0.171 \\
 & Beethoven & Schubert & -0.836 & 116.9 & 0.405 \\
\midrule
Quartet & Mozart & Haydn & \textbf{-6.367} & \textbf{44.8} & $\mathbf{9.08\times10^{-8}}$ \\
 & Mozart & Beethoven & 1.468 & 100.7 & 0.145 \\
 & Haydn & Beethoven & \textbf{6.870} & \textbf{83.4} & $\mathbf{1.07\times10^{-9}}$ \\
\bottomrule
\end{tabular}
\end{table}

\begin{table}[h!]
\centering
\caption{\textbf{Pairwise comparisons of the mean $\beta_1$ values obtained from an MTD-2 model.} Welch’s $t-$tests comparing composers in the sonata and quartet datasets. We report the $t$-statistics, degrees of freedom (df), and $p$-values for each test. The sign of the $t$-statistic indicates the direction: positive means the first-listed composer has the higher mean value. Boldface indicates significance at the Bonferroni-adjusted threshold ($p < 0.05/9$).}
\label{pair_beta1_MTD2}
\begin{tabular}{llllcc}
\toprule
Dataset & Composer 1 & Composer 2 & $t$-statistic & df & $p$-value \\
\midrule
Sonata & Mozart & Haydn & 0.926 & 144.1 & 0.356 \\
 & Mozart & Beethoven & \textbf{3.528} & \textbf{90.1} & $\mathbf{6.61\times10^{-4}}$ \\
 & Mozart & Schubert & 2.347 & 95.9 & 0.021 \\
 & Haydn & Beethoven & \textbf{2.861} & \textbf{100.9} & \textbf{0.005} \\
 & Haydn & Schubert & 1.569 & 119.2 & 0.119 \\
 & Beethoven & Schubert & -1.380 & 112.4 & 0.170 \\
\midrule
Quartet & Mozart & Haydn & \textbf{5.075} & \textbf{140.5} & $\mathbf{1.21\times10^{-6}}$ \\
 & Mozart & Beethoven & \textbf{7.008} & \textbf{101.6} & $\mathbf{2.71\times10^{-10}}$ \\
 & Haydn & Beethoven & 1.640 & 178.4 & 0.103 \\
\bottomrule
\end{tabular}
\end{table}

\begin{table}[h!]
\centering
\caption{\textbf{Pairwise comparisons of the mean $\beta_1$ values obtained from an MTD-3 model.} Welch’s $t-$tests comparing composers in the sonata and quartet datasets. We report the $t$-statistics, degrees of freedom (df), and $p$-values for each test. The sign of the $t$-statistic indicates the direction: positive means the first-listed composer has the higher mean value. Boldface indicates significance at the Bonferroni-adjusted threshold ($p < 0.05/9$).}
\label{pair_beta1_MTD3}
\begin{tabular}{llllcc}
\toprule
Dataset & Composer 1 & Composer 2 & $t$-statistic & df & $p$-value \\
\midrule
Sonata & Mozart & Haydn & 1.035 & 138.8 & 0.302 \\
 & Mozart & Beethoven & \textbf{4.065} & \textbf{91.0} & $\mathbf{1.02\times10^{-4}}$ \\
 & Mozart & Schubert & 2.673 & 94.0 & 0.009 \\
 & Haydn & Beethoven & \textbf{3.346} & \textbf{99.9} & \textbf{0.001} \\
 & Haydn & Schubert & 1.865 & 109.9 & 0.065 \\
 & Beethoven & Schubert & -1.478 & 114.9 & 0.142 \\
\midrule
Quartet & Mozart & Haydn & \textbf{4.759} & \textbf{113.2} & $\mathbf{5.79\times10^{-6}}$ \\
 & Mozart & Beethoven & \textbf{6.716} & \textbf{98.9} & $\mathbf{1.19\times10^{-9}}$ \\
 & Haydn & Beethoven & 2.189 & 164.0 & 0.030 \\
\bottomrule
\end{tabular}
\end{table}

\begin{table}[h!]
\centering
\caption{\textbf{Pairwise comparisons of the mean $\beta_2$ values obtained from an MTD-3 model.} Welch’s $t-$tests comparing composers in the sonata and quartet datasets. We report the $t$-statistics, degrees of freedom (df), and $p$-values for each test. The sign of the $t$-statistic indicates the direction: positive means the first-listed composer has the higher mean value. Boldface indicates significance at the Bonferroni-adjusted threshold ($p < 0.05/9$).}
\label{pair_beta2_mtd3}
\begin{tabular}{llllcc}
\toprule
Dataset & Composer 1 & Composer 2 & $t$-statistic & df & $p$-value \\
\midrule
Sonata & Mozart & Haydn & 0.933 & 75.0 & 0.354 \\
 & Mozart & Beethoven & -2.573 & 94.2 & 0.012 \\
 & Mozart & Schubert & -2.220 & 100.4 & 0.029 \\
 & Haydn & Beethoven & \textbf{-3.445} & \textbf{74.4} & $\mathbf
 {9.41\times10^{-4}}$ \\
 & Haydn & Schubert & \textbf{-3.389} & \textbf{82.7} & \textbf{0.001} \\
 & Beethoven & Schubert & 0.719 & 108.9 & 0.474 \\
\midrule
Quartet & Mozart & Haydn & -0.199 & 51.5 & 0.843 \\
 & Mozart & Beethoven & \textbf{-3.098} & \textbf{82.4} & \textbf{0.003} \\
 & Haydn & Beethoven & \textbf{-3.811} & \textbf{113.8} & $\mathbf{2.25\times10^{-4}}$ \\
\bottomrule
\end{tabular}
\end{table}

\begin{table}[h!]
\centering
\caption{\textbf{Pairwise comparisons of the mean $\beta_3$ values obtained from an MTD-3 model.} Welch’s $t-$tests comparing composers in the sonata and quartet datasets. We report the $t$-statistics, degrees of freedom (df), and $p$-values for each test. The sign of the $t$-statistic indicates the direction: positive means the first-listed composer has the higher mean value. Boldface indicates significance at the Bonferroni-adjusted threshold ($p < 0.05/9$).}
\label{pair_beta3_mtd3}
\begin{tabular}{llllcc}
\toprule
Dataset & Composer 1 & Composer 2 & $t$-statistic & df & $p$-value \\
\midrule
Sonata & Mozart & Haydn & -1.433 & 147.5 & 0.154 \\
 & Mozart & Beethoven & \textbf{-3.200} & \textbf{91.1} & \textbf{0.00}2 \\
 & Mozart & Schubert & -2.065 & 97.3 & 0.042 \\
 & Haydn & Beethoven & -2.148 & 104.8 & 0.034 \\
 & Haydn & Schubert & -0.829 & 127.0 & 0.409 \\
 & Beethoven & Schubert & 1.330 & 111.6 & 0.186 \\
\midrule
Quartet & Mozart & Haydn & \textbf{-5.289} & \textbf{138.7} & $\mathbf{4.68\times10^{-7}}$ \\
 & Mozart & Beethoven & \textbf{-6.433} & \textbf{101.0} & $\mathbf{4.22\times10^{-9}}$ \\
 & Haydn & Beethoven & -0.907 & 179.9 & 0.366 \\
\bottomrule
\end{tabular}
\end{table}

\clearpage
\subsection{Grouped comparisons (for a given composer vs. the other composers)} \label{sec:grouped_all}

In Sec.~\ref{sec:pairwise_all} we compared composers pairwise across metrics and noted that Mozart most consistently showed more localized statistical dependencies. Here, we test that hypothesis at the group level by comparing Mozart to the combined set of the other composers for each metric, separately for sonatas and quartets, using Welch’s $t$-tests with a Bonferroni-adjusted threshold $p < 0.05/12$. For consistency of direction across tables, we report the decay constant, the short-lag ratio, and the MTD lag-1 weights as their negatives (that is, $-\lambda$, $-I(1)/I(2)$, and $-\beta_1$ from the MTD-2 and MTD-3 models), so that smaller values of each metric indicate a stronger dependence on local structure, as opposed to higher-order correlations. Results are reported in Table~\ref{tab:Mozart_summary}. For completeness, we provide analogous grouped comparisons for the other composers in Tables~\ref{tab:Haydn_summary}–-\ref{tab:Schubert_summary}. 

For Mozart vs. all other composers (Table~\ref{tab:Mozart_summary}), the results are consistent across most metrics. For the sonatas, we observe significant differences for $-\lambda$, $-I(1)/I(2)$, $\Delta BIC$, and LLR, all indicating that Mozart's compositions rely to a lesser extent on higher-order dependencies than the other composers. Although the lag-1 weights ($-\beta_{1}$ from MTD-2 and MTD-3) do not reach the Bonferroni-adjusted significance threshold, the direction of the effect is consistent with the same interpretation. For the quartets, we observe significant differences for $-\lambda$, $-I(1)/I(2)$, LLR, and the MTD lag-1 weights, with the direction indicating stronger local dependence in Mozart. The effect in $\Delta$ BIC is weaker for quartets, a pattern consistent with our earlier observation in the pairwise comparisons (Table~\ref{pair_delta_bic}) that Mozart quartets do not differ substantially from Beethoven's quartets by this metric.

For the other composers (Tables~\ref{tab:Haydn_summary}–\ref{tab:Schubert_summary}), the results are more mixed and less consistent than for Mozart. For Haydn vs. the rest, significant differences are observed for $-\lambda$ and $\Delta BIC$ in both sonatas and quartets, with Haydn showing larger $-\lambda$ and more positive $\Delta BIC$. These outcomes suggest that Haydn’s pieces rely less strongly on immediate note-to-note structure and instead place relatively greater weight on longer-range dependencies. Additional differences are more limited: LLR is significantly lower only in quartets, and the MTD-based lag-1 weights do not differ significantly at the adjusted threshold. For Beethoven vs. the rest, the sonata results point toward weaker local dependence, with significantly larger values of $-I(1)/I(2)$ and $-\beta_{1}$ from both MTD-2 and MTD-3, while $\Delta$BIC does not differ significantly. The quartet results are more complex: Beethoven shows a more negative $\Delta$BIC (consistent with better fits from first-order models, in line with the pairwise results in Table~\ref{pair_delta_bic}), but at the same time significantly larger $-\beta_{1}$ values, which point to weaker local dependence. Finally, for Schubert vs. the rest, the only strong distinction arises in the sonatas, where Schubert exhibits significantly higher LLR values; other metrics do not yield consistent or significant contrasts. Overall, these findings indicate that, unlike Mozart, the other composers do not show a uniform shift toward or away from local statistical dependencies, but rather display patterns that vary depending on the metric and dataset considered.

\begin{table}[h!]
\centering
\caption{\textbf{Grouped comparisons of mean values for Mozart vs. all other composers.} 
Welch’s $t$-tests compare Mozart to the combined set of the other composers in the sonata and quartet datasets. For ease of interpretation, some metrics are shown as their negative values so that smaller values indicate stronger local dependence as opposed to longer-range dependence. We report group means, standard deviations, $N$, and Welch’s $t$-test results. Boldface indicates significance at the Bonferroni-adjusted threshold ($p < 0.05/12$).}
\label{tab:Mozart_summary}
\begin{tabular}{llcccc}
\toprule
Feature & Group & Mean & Std. Dev. & N & $t$-test (Mozart vs Others) \\
\midrule
\multicolumn{6}{l}{\textbf{Negative Decay Constant}} \\
\midrule
Sonata & Mozart & -0.2477 & 0.1112 & 45 & \multirow{2}{*}{$\mathbf{t = -6.07,\, p = 1.83\!\times\!10^{-7},\, df = 48.94}$} \\
       & Others (H+B+S) & -0.1432 & 0.0654 & 278 & \\
Quartet & Mozart & -0.2337 & 0.1235 & 34 & \multirow{2}{*}{$\mathbf{t = -3.09,\, p = 0.004,\, df = 37.13}$} \\
       & Others (H+B) & -0.1653 & 0.0738 & 194 & \\
\midrule
\multicolumn{6}{l}{\textbf{Negative I(1)/I(2)}} \\
\midrule
Sonata & Mozart & -1.8283 & 0.3915 & 45 & \multirow{2}{*}{$\mathbf{t = -6.24,\, p = 5.85\!\times\!10^{-8},\, df = 57.03}$} \\
       & Others (H+B+S) & -1.4352 & 0.3679 & 278 & \\
Quartet & Mozart & -1.7116 & 0.2885 & 34 & \multirow{2}{*}{$\mathbf{t = -3.51,\, p = 9.61\!\times\!10^{-4},\, df = 49.09}$} \\
       & Others (H+B) & -1.5163 & 0.3309 & 194 & \\
\midrule
\multicolumn{6}{l}{\textbf{LLR}} \\
\midrule
Sonata & Mozart & 1263.0777 & 610.9288 & 45 & \multirow{2}{*}{$\mathbf{t = -4.65,\, p = 1.16\!\times\!10^{-5},\, df = 88.10}$} \\
       & Others (H+B+S) & 1776.4308 & 1012.4792 & 278 & \\
Quartet & Mozart & 1008.3823 & 479.3164 & 34 & \multirow{2}{*}{$\mathbf{t = -3.76,\, p = 3.47\!\times\!10^{-4},\, df = 69.82}$} \\
       & Others (H+B) & 1390.6869 & 805.8790 & 194 & \\
\midrule
\multicolumn{6}{l}{\textbf{BIC1-BIC2}} \\
\midrule
Sonata & Mozart & -751.7989 & 755.5530 & 45 & \multirow{2}{*}{$\mathbf{t = -6.94,\, p = 4.86\!\times\!10^{-9},\, df = 54.82}$} \\
       & Others (H+B+S) & 83.4394 & 649.5142 & 278 & \\
Quartet & Mozart & -405.4947 & 431.0613 & 34 & \multirow{2}{*}{$\mathbf{t = -3.01,\, p = 0.004,\, df = 61.75}$} \\
       & Others (H+B) & -140.0453 & 646.9220 & 194 & \\
\midrule
\multicolumn{6}{l}{\textbf{Negative $\beta_1$ (MTD-2)}} \\
\midrule
Sonata & Mozart & -0.8359 & 0.1214 & 45 & \multirow{2}{*}{$t = -2.54,\, p = 0.012,\, df = 109.43$} \\
       & Others (H+B+S) & -0.7766 & 0.2412 & 279 & \\
Quartet & Mozart & -0.9314 & 0.0713 & 34 & \multirow{2}{*}{$\mathbf{t = -6.87,\, p = 3.53\!\times\!10^{-10},\, df = 113.80}$} \\
       & Others (H+B) & -0.8103 & 0.1735 & 194 & \\
\midrule
\multicolumn{6}{l}{\textbf{Negative $\beta_1$ (MTD-3)}} \\
\midrule
Sonata & Mozart & -0.8087 & 0.1275 & 45 & \multirow{2}{*}{$t = -2.90,\, p = 0.005,\, df = 107.89$} \\
       & Others (H+B+S) & -0.7379 & 0.2507 & 279 & \\
Quartet & Mozart & -0.9114 & 0.0879 & 34 & \multirow{2}{*}{$\mathbf{t = -6.39,\, p = 8.17\!\times\!10^{-9},\, df = 86.04}$} \\
       & Others (H+B) & -0.7847 & 0.1749 & 194 & \\
\midrule
\bottomrule
\end{tabular}
\end{table}

\begin{table}[h!]
\centering
\caption{\textbf{Grouped comparisons of mean values for Haydn vs. all other composers.} 
Welch’s $t$-tests compare Haydn to the combined set of the other composers in the sonata and quartet datasets. For ease of interpretation, some metrics are shown as their negative values so that smaller values indicate stronger local dependence as opposed to longer-range dependence. We report group means, standard deviations, $N$, and Welch’s $t$-test results. Boldface indicates significance at the Bonferroni-adjusted threshold ($p < 0.05/12$).}
\label{tab:Haydn_summary}
\begin{tabular}{llcccc}
\toprule
Feature & Group & Mean & Std. Dev. & N & $t$-test (Haydn vs Others) \\
\midrule
\multicolumn{6}{l}{\textbf{Negative Decay Constant}} \\
\midrule
Sonata & Haydn & -0.1329 & 0.0612 & 159 & \multirow{2}{*}{$\mathbf{t = 5.64,\, p = 4.02\!\times\!10^{-8},\, df = 285.01}$} \\
       & Others (M+B+S) & -0.1819 & 0.0917 & 164 & \\
Quartet & Haydn & -0.1591 & 0.0704 & 124 & \multirow{2}{*}{$\mathbf{t = 3.08,\, p = 0.002,\, df = 181.42}$} \\
       & Others (M+B) & -0.1950 & 0.0991 & 104 & \\
\midrule
\multicolumn{6}{l}{\textbf{Negative I(1)/I(2)}} \\
\midrule
Sonata & Haydn & -1.4706 & 0.3755 & 159 & \multirow{2}{*}{$t = 0.86,\, p = 0.388,\, df = 319.69$} \\
       & Others (M+B+S) & -1.5087 & 0.4130 & 164 & \\
Quartet & Haydn & -1.5049 & 0.3208 & 124 & \multirow{2}{*}{$t = 2.01,\, p = 0.046,\, df = 214.39$} \\
       & Others (M+B) & -1.5938 & 0.3392 & 104 & \\
\midrule
\multicolumn{6}{l}{\textbf{LLR}} \\
\midrule
Sonata & Haydn & 1555.4273 & 757.6133 & 159 & \multirow{2}{*}{$t = -2.73,\, p = 0.007,\, df = 284.24$} \\
       & Others (M+B+S) & 1849.8373 & 1141.5757 & 164 & \\
Quartet & Haydn & 1151.5987 & 623.9934 & 124 & \multirow{2}{*}{$\mathbf{t = -3.86,\, p = 1.59\!\times\!10^{-4},\, df = 180.96}$} \\
       & Others (M+B) & 1550.7694 & 881.0897 & 104 & \\
\midrule
\multicolumn{6}{l}{\textbf{BIC1-BIC2}} \\
\midrule
Sonata & Haydn & 172.3607 & 524.3473 & 159 & \multirow{2}{*}{$\mathbf{t = 5.23,\, p = 3.31\!\times\!10^{-7},\, df = 276.43}$} \\
       & Others (M+B+S) & -231.9521 & 830.3511 & 164 & \\
Quartet & Haydn & 111.1118 & 342.4448 & 124 & \multirow{2}{*}{$\mathbf{t = 8.36,\, p = 5.02\!\times\!10^{-14},\, df = 142.80}$} \\
       & Others (M+B) & -526.2832 & 707.2843 & 104 & \\
\midrule
\multicolumn{6}{l}{\textbf{Negative $\beta_1$ (MTD-2)}} \\
\midrule
Sonata & Haydn & -0.8114 & 0.2400 & 160 & \multirow{2}{*}{$t = -2.06,\, p = 0.040,\, df = 316.34$} \\
       & Others (M+B+S) & -0.7590 & 0.2151 & 164 & \\
Quartet & Haydn & -0.8245 & 0.1888 & 124 & \multirow{2}{*}{$t = 0.39,\, p = 0.694,\, df = 222.54$} \\
       & Others (M+B) & -0.8331 & 0.1392 & 104 & \\
\midrule
\multicolumn{6}{l}{\textbf{Negative $\beta_1$ (MTD-3)}} \\
\midrule
Sonata & Haydn & -0.7805 & 0.2439 & 160 & \multirow{2}{*}{$t = -2.45,\, p = 0.015,\, df = 319.56$} \\
       & Others (M+B+S) & -0.7158 & 0.2290 & 164 & \\
Quartet & Haydn & -0.8043 & 0.1829 & 124 & \multirow{2}{*}{$t = -0.07,\, p = 0.946,\, df = 225.95$} \\
       & Others (M+B) & -0.8027 & 0.1555 & 104 & \\
\midrule
\bottomrule
\end{tabular}
\end{table}

\begin{table}[h!]
\centering
\caption{\textbf{Grouped comparisons of mean values for Beethoven vs. all other composers.} 
Welch’s $t$-tests compare Beethoven to the combined set of the other composers in the sonata and quartet datasets. For ease of interpretation, some metrics are shown as their negative values so that smaller values indicate stronger local dependence as opposed to longer-range dependence. We report group means, standard deviations, $N$, and Welch’s $t$-test results. Boldface indicates significance at the Bonferroni-adjusted threshold ($p < 0.05/12$).}
\label{tab:Beethoven_summary}
\begin{tabular}{llcccc}
\toprule
Feature & Group & Mean & Std. Dev. & N & $t$-test (Beethoven vs Others) \\
\midrule
\multicolumn{6}{l}{\textbf{Negative Decay Constant}} \\
\midrule
Sonata & Beethoven & -0.1446 & 0.0688 & 61 & \multirow{2}{*}{$t = 1.57,\, p = 0.119,\, df = 105.73$} \\
       & Others (M+H+S) & -0.1608 & 0.0844 & 262 & \\
Quartet & Beethoven & -0.1762 & 0.0781 & 70 & \multirow{2}{*}{$t = -0.09,\, p = 0.929,\, df = 150.63$} \\
       & Others (M+H) & -0.1751 & 0.0901 & 158 & \\
\midrule
\multicolumn{6}{l}{\textbf{Negative I(1)/I(2)}} \\
\midrule
Sonata & Beethoven & -1.3463 & 0.3450 & 61 & \multirow{2}{*}{$\mathbf{t = 3.48,\, p = 7.49\!\times\!10^{-4},\, df = 100.37}$} \\
       & Others (M+H+S) & -1.5234 & 0.3990 & 262 & \\
Quartet & Beethoven & -1.5365 & 0.3471 & 70 & \multirow{2}{*}{$t = 0.26,\, p = 0.794,\, df = 124.46$} \\
       & Others (M+H) & -1.5494 & 0.3254 & 158 & \\
\midrule
\multicolumn{6}{l}{\textbf{LLR}} \\
\midrule
Sonata & Beethoven & 1835.7685 & 1137.8560 & 61 & \multirow{2}{*}{$t = 1.02,\, p = 0.310,\, df = 79.87$} \\
       & Others (M+H+S) & 1674.4442 & 940.4371 & 262 & \\
Quartet & Beethoven & 1814.2146 & 910.7720 & 70 & \multirow{2}{*}{$\mathbf{t = 5.80,\, p = 8.53\!\times\!10^{-8},\, df = 96.17}$} \\
       & Others (M+H) & 1120.7800 & 598.7340 & 158 & \\
\midrule
\multicolumn{6}{l}{\textbf{BIC1-BIC2}} \\
\midrule
Sonata & Beethoven & -93.1846 & 795.3588 & 61 & \multirow{2}{*}{$t = -0.67,\, p = 0.508,\, df = 83.18$} \\
       & Others (M+H+S) & -18.8951 & 707.4473 & 262 & \\
Quartet & Beethoven & -584.9520 & 801.5293 & 70 & \multirow{2}{*}{$\mathbf{t = -5.72,\, p = 1.48\!\times\!10^{-7},\, df = 86.17}$} \\
       & Others (M+H) & -0.0567 & 420.8214 & 158 & \\
\midrule
\multicolumn{6}{l}{\textbf{Negative $\beta_1$ (MTD-2)}} \\
\midrule
Sonata & Beethoven & -0.7011 & 0.2599 & 61 & \multirow{2}{*}{$t = 2.86,\, p = 0.005,\, df = 80.22$} \\
       & Others (M+H+S) & -0.8043 & 0.2170 & 263 & \\
Quartet & Beethoven & -0.7853 & 0.1391 & 70 & \multirow{2}{*}{$t = 2.84,\, p = 0.005,\, df = 164.53$} \\
       & Others (M+H) & -0.8475 & 0.1760 & 158 & \\
\midrule
\multicolumn{6}{l}{\textbf{Negative $\beta_1$ (MTD-3)}} \\
\midrule
Sonata & Beethoven & -0.6481 & 0.2674 & 61 & \multirow{2}{*}{$\mathbf{t = 3.30,\, p = 0.001,\, df = 80.60}$} \\
       & Others (M+H+S) & -0.7709 & 0.2253 & 263 & \\
Quartet & Beethoven & -0.7500 & 0.1538 & 70 & \multirow{2}{*}{$\mathbf{t = 3.35,\, p = 0.001,\, df = 146.86}$} \\
       & Others (M+H) & -0.8273 & 0.1728 & 158 & \\
\midrule
\bottomrule
\end{tabular}
\end{table}

\begin{table}[h!]
\centering
\caption{\textbf{Grouped comparisons of mean values for Schubert vs. all other composers.} 
Welch’s $t$-tests compare Schubert to the combined set of the other composers in the sonata and quartet datasets. For ease of interpretation, some metrics are shown as their negative values so that smaller values indicate stronger local dependence as opposed to longer-range dependence. We report group means, standard deviations, $N$, and Welch’s $t$-test results. Boldface indicates significance at the Bonferroni-adjusted threshold ($p < 0.05/12$).}
\label{tab:Schubert_summary}
\begin{tabular}{llcccc}
\toprule
Feature & Group & Mean & Std. Dev. & N & $t$-test (Schubert vs Others) \\
\midrule
\multicolumn{6}{l}{\textbf{Negative Decay Constant}} \\
\midrule
Sonata & Schubert & -0.1699 & 0.0651 & 58 & \multirow{2}{*}{$t = -1.47,\, p = 0.144,\, df = 103.54$} \\
       & Others (M+H+B) & -0.1551 & 0.0849 & 265 & \\
\midrule
\multicolumn{6}{l}{\textbf{Negative I(1)/I(2)}} \\
\midrule
Sonata & Schubert & -1.4315 & 0.3540 & 58 & \multirow{2}{*}{$t = 1.34,\, p = 0.183,\, df = 91.78$} \\
       & Others (M+H+B) & -1.5027 & 0.4029 & 265 & \\
\midrule
\multicolumn{6}{l}{\textbf{LLR}} \\
\midrule
Sonata & Schubert & 2319.8783 & 1243.1283 & 58 & \multirow{2}{*}{$\mathbf{t = 4.33,\, p = 4.86\!\times\!10^{-5},\, df = 69.20}$} \\
       & Others (M+H+B) & 1570.3144 & 859.1727 & 265 & \\
\midrule
\multicolumn{6}{l}{\textbf{BIC1-BIC2}} \\
\midrule
Sonata & Schubert & 25.4321 & 738.3623 & 58 & \multirow{2}{*}{$t = 0.66,\, p = 0.510,\, df = 82.21$} \\
       & Others (M+H+B) & -45.6975 & 721.9640 & 265 & \\
\midrule
\multicolumn{6}{l}{\textbf{Negative $\beta_1$ (MTD-2)}} \\
\midrule
Sonata & Schubert & -0.7602 & 0.2007 & 58 & \multirow{2}{*}{$t = 1.00,\, p = 0.322,\, df = 93.72$} \\
       & Others (M+H+B) & -0.7903 & 0.2347 & 266 & \\
\midrule
\multicolumn{6}{l}{\textbf{Negative $\beta_1$ (MTD-3)}} \\
\midrule
Sonata & Schubert & -0.7150 & 0.2213 & 58 & \multirow{2}{*}{$t = 1.21,\, p = 0.228,\, df = 88.74$} \\
       & Others (M+H+B) & -0.7549 & 0.2417 & 266 & \\
\midrule
\bottomrule
\end{tabular}
\end{table}

\end{document}